\documentclass[aps,preprint,showpacs,preprintnumbers,amsmath,amssymb]{revtex4}
\usepackage{graphicx}
\pdfoutput=1
\def\beq{\begin{equation}}
\def\eeq{\end{equation}}
\def\beqn{\begin{eqnarray}}
\def\eeqn{\end{eqnarray}}

\newcounter{saveeqn}

\def\r {{\bf r}}
\def\R {{\bf R}}
\def\s {{\bf s}}
\def\p {{\bf k}}

\def\x {{x}}

\def\k {{k}}

\newcommand{\promille}{%
  \relax\ifmmode\promillezeichen
        \else\leavevmode\(\mathsurround=0pt\promillezeichen\)\fi}
\newcommand{\promillezeichen}{%
  \kern-.05em%
  \raise.5ex\hbox{\the\scriptfont0 0}%
  \kern-.15em/\kern-.15em%
  \lower.25ex\hbox{\the\scriptfont0 00}}

\begin{document}
\title{Reduced density matrices and coherence of trapped interacting bosons}
\author{Kaspar \ Sakmann\footnote{Corresponding author, 
E-mail: kaspar.sakmann@pci.uni-heidelberg.de}, Alexej I.\ Streltsov\footnote{E-mail: alexej.streltsov@pci.uni-heidelberg.de},  Ofir E. Alon\footnote{E-mail: ofir.alon@pci.uni-heidelberg.de},
and Lorenz S.\ Cederbaum\footnote{E-mail: lorenz.cederbaum@pci.uni-heidelberg.de}}

\affiliation{Theoretische Chemie, Universit\"{a}t Heidelberg, D-69120 Heidelberg, Germany}


\begin{abstract}
The first- and second-order 
correlation functions of trapped, interacting Bose-Einstein condensates
are investigated numerically 
on a many-body level from first principles. 
Correlations in real space and  momentum space are treated.
The coherence properties are analyzed.
The results are obtained by solving the 
many-body Schr\"odinger equation.
It is shown in an example how many-body effects can 
be induced by the trap geometry. 
A generic fragmentation scenario of a condensate is 
considered. The correlation functions
are discussed along a pathway from
a single condensate to a fragmented condensate.
It is shown that strong correlations can arise 
from the geometry of the trap, 
even at weak interaction strengths.
The natural orbitals and natural geminals of 
the system are obtained and discussed. 
It is shown how the fragmentation of the condensate 
can be understood in terms of its natural geminals.
The many-body results are compared to those
of mean-field theory. The best solution
within mean-field theory  is obtained.
The limits in which mean-field theories are valid
are determined. In these limits the behavior of the correlation functions 
is explained within an analytical model.
\end{abstract}

\maketitle

\section{INTRODUCTION}
The computation of correlation functions in interacting 
quantum many-body systems is a challenging problem of contemporary physics. 
Correlations between particles can exist in time, in real space or in momentum space.
Of course, all combinations of the above three cases are possible.
Since the first experimental realization of Bose-Einstein condensates (BECs) 
in ultracold atomic gases \cite{firstBEC1,firstBEC2,firstBEC3}, 
great experimental and theoretical 
progress has been made in the determination of the coherence 
and the correlation functions of Bose-condensed systems. Over the years
experiments have measured more and more accurately first, second and to some extent even third
order correlations of trapped BECs, see \cite{BGM97,CHK03,LHP04,FGW05,SHP05,ORK05,SHL07,ROD07}.
Theoretically, the correlation functions of trapped, interacting 
BECs have been investigated in numerous works, 
see e.g.  \cite{NaG99,HoC99,AsG03,KGD03,BaR04,GPS06,ZMS06a,ZMS06b,PeB07,ZMS08}.
While analytical approaches from first principles 
are usually restricted to treat homogeneous gases without any 
trapping potential, numerical methods can overcome this restriction.  
It is important to note that the shape of the trapping potential
can have a substantial impact on the properties of the many-body system.
This is particularly true for issues concerning
condensation \cite{PeO56} and fragmentation of Bose systems \cite{NoS82}.
For example, the ground state of weakly interacting 
condensates in harmonic traps is almost fully 
condensed,  while the ground state
of double-well potentials can be 
fragmented or condensed, depending 
on the height of the barrier, the number of particles and the 
interaction strength \cite{SpS99,CeS03,SCM04,AlC05,ASC05b,SAC06}.
In this work we investigate  
first- and second-order correlations
of trapped, interacting condensates and their coherence properties 
depending on the trap geometry from first principles. 
Our results are obtained by solving the 
many-body Schr\"odinger equation of the interacting system numerically. 
From this many-body solution 
we extract the first- and second-order reduced density matrices 
which allow us to compute all real and momentum space 
first- and second-order correlations and in particular 
the fragmentation of the condensate.
For illustration purposes we consider a 
stationary system in the ground state to 
show how many-body effects can become dominant when the trap geometry is varied.
As a numerical method to solve the interacting many-body problem  
we use the recently developed 
{\em multiconfigurational time-dependent Hartree for bosons} (MCTDHB), 
which propagates a given many-body state in time \cite{SAC07,ASC07b,ASC07c}. 
By propagation in imaginary time it allows us to investigate 
the ground state and other stationary states. 
Alternatively, one can use the stationary multiconfigurational
Hartree for bosons to compute these states \cite{SAC06}.
In order to identify true many-body effects in the correlation functions, 
we compare our many-body results with those based on  mean-field approaches. 
More specifically, we compute the energetically-lowest mean-field 
solution of the same system for comparison. 
This is the best approximation to the true many-body wave-function 
within mean-field theory and, thereby, allows us to pinpoint the limits of mean-field
theory. A general method to compute this {\em best mean-field} solution has been 
developed in our group \cite{CeS03,SCM04,ASC05a,ASC05b,AlC05}.
For completeness we compare the many-body results also to the widely 
used Gross-Pitaevskii mean-field solution.
In order to understand first- and 
second-order correlations in an intuitive way, 
we develop an analytical mean-field model 
which explains the general structure of our results
in those regions where many-body effects can be 
safely neglected.

This paper is organized as follows. In Sec.~\ref{secbasic} we review
basic facts about reduced density matrices, correlation functions and coherence.
In Sec.~\ref{sectheory} we give a brief introduction to the 
numerical many-body and mean-field 
methods that we use. In Sec.~\ref{model} we introduce a generic 
one-dimensional model system that we solve. 
In particular, we identify mean-field and many-body 
regimes of this system. We explain the transition from a condensed to a fragmented 
state in terms of the natural geminals of the system.
In Secs.~\ref{firstOrderCorrelations} and \ref{secondOrderCorrelations} 
we present our many-body results for the first- and second-order correlation
functions and  the coherence properties of the model system. We compare our many-body results with 
those obtained by using the  {\em best mean-field} and the Gross-Pitaevskii approximation. 
The structure of the correlation functions 
and the coherence are explained within an analytical mean-field model in the limits 
where mean-field theory is applicable.

\section{BASIC DEFINITIONS}\label{secbasic}
We consider a given wave function $\Psi(\r_1,\dots,\r_N;t)$ of $N$ identical,  
spinless bosons with spatial coordinates $\r_i$ in $D$ dimensions.
The $p$-th order reduced density matrix (RDM), is defined by
\begin{eqnarray}\label{DEFRHOP}
\rho^{(p)}(\r_1,\dots,\r_p\vert \r'_1,\dots,\r'_p;t)=
\frac{N!}{(N-p)!}\int\Psi(\r_1,\dots,\r_p,\r_{p+1},\dots,\r_N;t)\nonumber\\
\times \Psi^\ast(\r'_1,\dots,\r'_p,\r_{p+1},\dots,\r_N;t) d\r_{p+1}\dots d\r_{N},
\end{eqnarray}
where the wave function is assumed to be normalized $\langle\Psi(t)\vert\Psi(t)\rangle=1$.
Equivalent to Eq.~(\ref{DEFRHOP}), $\rho^{(p)}(\r_1,\dots,\r_p\vert \r'_1,\dots,\r'_p;t)$ 
can be regarded as the kernel of the operator 
\begin{equation} 
\hat{\rho}^{(p)}=\frac{N!}{(N-p)!}{\text{Tr}}_{N-p}\left[\vert \Psi(t)\rangle\langle\Psi(t)\vert\right]
\end{equation}
in Hilbert space, where ${\text {Tr}}_{N-p}[\cdot]$ specifies taking the partial trace over
$N-p$ particles. Since the wave function is symmetric in its coordinates, it does not matter over which particles 
the trace is taken.  In what follows, we add  
$\vert\Psi\rangle$ as an additional 
subscript if a result is only 
valid for states $\vert\Psi\rangle$ 
of a particular form.

The diagonal $\rho^{(p)}(\r_1,\dots,\r_p\vert \r_1,\dots,\r_p;t)$ 
is the $p$-particle probability distribution at time $t$ multiplied by $N!/(N-p)!$.
The $p$-th order RDM $\rho^{(p)}$ can be expanded in its eigenfunctions, leading to the representation
\begin{eqnarray}\label{NATURALREP}
\rho^{(p)}(\r_1,\dots,\r_p\vert \r'_1,\dots,\r'_p;t)=\sum_i n^{(p)}_i(t)\alpha^{(p)}_i(\r_1,\dots,\r_p,t){\alpha^{(p)}_i}^\ast(\r'_1,\dots,\r'_p,t).
\end{eqnarray}
Here, $n^{(p)}_{i}(t)$ denotes the $i$-th eigenvalue of the 
$p$-th order RDM and $\alpha^{(p)}_i(\r_1,\dots,\r_p,t)$ the corresponding
eigenfunction.
The eigenfunctions are known as {\em natural $p$-functions}
and the eigenvalues as {\em natural occupations}.
For $p=1$ and $p=2$ the eigenfunctions are also known as
{\em natural orbitals} and {\em natural geminals}, respectively.
We order the eigenvalues $n^{(p)}_{i}(t)$
for every $p$ non-increasingly, such that $n^{(p)}_{1}(t)$
denotes the largest eigenvalue of the $p$-th order RDM.
The normalization of the many-body 
wave function and Eqs.~(\ref{DEFRHOP}) and (\ref{NATURALREP})
put the restriction
\begin{eqnarray}\label{resnip}
\sum_{i} n_i^{(p)}(t)=\frac{N!}{(N-p)!}
\end{eqnarray}
on the eigenvalues of the $p$-th order RDM. 
Thus the largest eigenvalue $n^{(p)}_1(t)$ 
is bounded from above by \cite{CoY00,Maz07}
\begin{equation}\label{upperbound} 
n^{(p)}_{1}(t)\le\frac{N!}{(N-p)!}.
\end{equation}
Lower bounds on $n^{(p)}_{1}(t)$  can be derived, 
relating RDMs of different order \cite{Yan62,Sas65}. 
In particular, for the case $p=2$
it can be shown that \cite{Yan62} 
\begin{equation}\label{lowerbound}
n^{(2)}_{1}(t)\ge n^{(1)}_{1}(t)[n^{(1)}_{1}(t)-1].
\end{equation}

It is a well known fact that the natural 
orbitals of a symmetric (or antisymmetric) function $\Psi$ 
constitute a sufficient one-particle basis
to expand $\Psi$ and 
the eigenfunctions of the RDMs
for all $p$ \cite{CoY00}. 
It is therefore  
possible to construct the natural geminals 
in the basis of one-particle functions
spanned by the natural orbitals.

The determination of accurate bounds on 
eigenvalues of RDMs
is an active field of research \cite{CoY00,Maz07} 
since it is possible to express the {\em exact} 
energy expectation value of a quantum 
system of identical particles 
interacting via two-body interactions
by an expression involving only the 
natural geminals, $\alpha_i^{(2)}(\r_1,\r_2,t)$, 
and their occupations, $n_i^{(2)}(t)$. 
For a general Hamiltonian  
\begin{equation}\label{genHam}
H=\sum_{i=1}^N h(\r_i)+\sum_{i<j}^N W(\r_i-\r_j),
\end{equation}
consisting of one-body operators $h(\r_i)$ and 
two-body operators $W(\r_i-\r_j)$, the expectation value 
of the energy $E$ can be expressed by 
making use of the time-dependent natural 
geminals $\alpha_i^{(2)}(\x_1,\x_2,t)$ and following Refs. \cite{CoY00,Maz07} through
the equation:
\begin{equation}\label{gemEnergy}
E=\frac{1}{2}\sum_i n_i^{(2)}(t)\int d\r_1 d\r_2 \alpha^{{(2)}^\ast}_i(\r_1,\r_2,t)
\left[\frac{h(\r_1)+h(\r_2)}{N-1}+W(\r_1-\r_2) \right]
\alpha^{(2)}_i(\r_1,\r_2,t).
\end{equation} 
Note that the many-body wave function does not 
appear explicitly in Eq.~(\ref{gemEnergy}). 
We will not go any further into the details 
of these approaches to many-body physics
and refer the reader to the literature \cite{CoY00,Maz07}.

The natural orbitals also serve to define 
Bose-Einstein condensation in interacting 
systems. According to Penrose and Onsager \cite{PeO56},
a system of identical bosons is said to be {\em condensed}, 
if the largest eigenvalue of the 
first-order RDM is of the order of the 
number of particles in the system, $n^{(1)}_1={\mathcal O}(N)$.
If more than one eigenvalue of the first-order RDM 
is of the order of the number of particles, 
the condensate is said to be {\em fragmented}, 
according to Nozi\`eres and  Saint James \cite{NoS82}.

Equivalent to Eq.~(\ref{DEFRHOP}), the $p$-th order RDM can be expressed through 
field operators as 
\begin{eqnarray}\label{DEFRHOP2}
\rho^{(p)}(\r_1,\dots,\r_p\vert \r'_1,\dots,\r'_p;t)=
\langle\Psi(t)\vert \hat{\mathbf \Psi}^\dag(\r'_1)\dots\hat{\mathbf \Psi}^\dag(\r'_p) 
\hat{\mathbf \Psi}(\r_p)\dots
\hat{\mathbf \Psi}(\r_1)\vert \Psi(t)\rangle,
\end{eqnarray}
where the Schr\"odinger field operators satisfy the usual 
bosonic commutation relations
\begin{equation}\label{com_rel}
\left[\hat{\mathbf \Psi}(\r),\hat{\mathbf \Psi}^\dag(\r')\right] = \delta(\r-\r'), \,\,\,\,\,\,\,
\left[\hat{\mathbf \Psi}(\r),\hat{\mathbf \Psi}(\r')\right] = 0.
\end{equation}
The representation given in Eq.~(\ref{DEFRHOP2}) shows that the $p$-th order RDM is identical to
the $p$-th order correlation function at equal times \cite{Gla63,NaG99}. 

In order to discuss correlations not only in real space,  but also in momentum space, 
we define the Fourier transform  of a function $f(\r_1,\dots,\r_p)$ of $p$  $D$-dimensional 
coordinates $\r_i$ by
\begin{equation}\label{DEFFT} 
f(\p_1,\dots,\p_p)=\frac{1}{(2\pi)^{pD/2}}\int d^p\r\, e^{-i\sum_{l=1}^p \p_l\r_l}f(\r_1,\dots,\r_p).
\end{equation} 
By applying the Fourier transform, Eq.~(\ref{DEFFT}), 
to the coordinates $\r_1,\dots,\r_p$ and $\r'_1,\dots,\r'_p$ of the natural $p$-functions 
$\alpha_i^{(p)}(\r_1,\dots,\r_p,t)$ and $\alpha_i^{(p)}(\r'_1,\dots,\r'_p,t)$ 
in Eq.~(\ref{NATURALREP}),
one arrives at the momentum space representation of $\hat{\rho}^{(p)}$: 
\begin{eqnarray}\label{MOMENTUMRHOP}
\rho^{(p)}(\p_1,\dots,\p_p\vert \p'_1,\dots,\p'_p;t) &=& \sum_i n^{(p)}_{i}(t)\alpha^{(p)}_i(\p_1,\dots,\p_p,t){\alpha^{(p)}_i}^\ast(\p_1',\dots,\p_p',t).
\end{eqnarray}
The diagonal $\rho^{(p)}(\p_1,\dots,\p_p\vert \p_1,\dots,\p_p;t)$ 
in momentum space is the $p$-particle 
momentum distribution at time $t$, 
multiplied by $N!/(N-p)!$. 
It can be shown that the $p$-particle
momentum distribution at large momenta is dominated by
contributions of $\rho^{(p)}(\r_1,\dots,\r_p\vert \r'_1,\dots,\r'_p;t)$ 
close to the diagonal, i.e. $\r_i\approx\r'_i$ for $i=1,\dots,p$.
Similarly, the $p$-particle distribution at 
low momenta is dominated by the behavior of 
$\rho^{(p)}(\r_1,\dots,\r_p\vert \r'_1,\dots,\r'_p;t)$ 
on the off-diagonal at large distances between $\r_i$ and $\r_i'$. 
See Appendix \ref{A} for more details.
	
Apart from the $p$-particle distributions themselves, 
either in real space or in momentum space, 
it is also of great interest to compare 
the $p$-particle probabilities to 
their respective one-particle probabilities. 
Thereby, it becomes possible to
identify effects that are due to 
the quantum statistics of the particles.
The {\em normalized $p$-th order correlation function} 
at time $t$ is defined by \cite{Gla63}
\begin{eqnarray}\label{gp}
g^{(p)}(\r'_1,\dots,\r'_p,\r_1,\dots,\r_p;t)&=&\frac{\rho^{(p)}(\r_1,\dots,\r_p\vert \r'_1,\dots,\r'_p;t)}
{\sqrt{\prod_{i=1}^p\rho^{(1)}(\r_i\vert\r_i;t)\rho^{(1)}(\r'_i\vert\r'_i;t)}}
\end{eqnarray}
and is the key  quantity in the definition of spatial coherence.
Full spatial $p$-th order coherence is obtained if   
$\rho^{(n)}(\r_1,\dots\r_p\vert\r'_1,\dots,\r'_p;t)$ 
factorizes for all $n\le p$ 
into a product of {\it one} complex valued function 
${\mathcal E(\r,t)}$ of the form 
\begin{equation}\label{coh}
\rho^{(n)}(\r_1,\dots,\r_p\vert\r'_1,\dots,\r'_p;t)={\mathcal E}^\ast(\r'_1,t)\cdots{\mathcal E}^\ast(\r'_n,t){\mathcal E}(\r_n,t)\cdots{\mathcal E}(\r_1,t).
\end{equation}
In this case 
\begin{equation}\label{abscoh}
\vert g^{(n)}(\r'_1,\dots,\r'_p,\r_1,\dots,\r_p;t)\vert=1
\end{equation}
for all $n\le p$. 
Otherwise, the state $\vert\Psi(t)\rangle$ is only 
partially coherent.
Full coherence in a system with a definite number
of particles $N$ can only be obtained for $p=1$ \cite{TiG65}.
However, when the particle number $N$ is large, $p$-th order coherence can be obtained 
up to corrections $\mathcal{O}(1/N)$, at least for $p\ll N$ \cite{TiG65}.

The diagonal of the normalized 
$p$-th order correlation function
$g^{(p)}(\r_1,\dots,\r_p,\r_1,\dots,\r_p;t)$ 
gives a measure for the degree of $p$-th order coherence.
For values $g^{(p)}(\r_1,\dots,\r_p,\r_1,\dots,\r_p;t)>1$ $(<1)$ the 
detection probabilities at positions 
$\r_1,\dots,\r_p$ are correlated (anticorrelated).

Note that if Eq.~(\ref{coh}) holds in real space, it must also hold in momentum space, 
as can be seen by Fourier transforming each of the $2n$ variables in Eq.~(\ref{coh}).
It is therefore possible to define 
the normalized $p$-th order correlation function in momentum space by
\begin{eqnarray}\label{gpk}
g^{(p)}(\p'_1,\dots,\p'_p,\p_1,\dots,\p_p;t)&=&
\frac{\rho^{(p)}(\p_1,\dots,\p_p\vert \p'_1,\dots,\p'_p;t)}
{\sqrt{\prod_{i=1}^p\rho^{(1)}(\p_i\vert \p_i;t)\rho^{(1)}(\p'_i\vert \p'_i;t)}}.
\end{eqnarray}
The diagonal of Eq.~(\ref{gpk}), $g^{(p)}(\p_1,\dots,\p_p,\p_1,\dots,\p_p;t)$, 
expresses  the tendency of $p$ momenta to be measured simultaneously.
For values $g^{(p)}(\p_1,\dots,\p_p,\p_1,\dots,\p_p;t)>1$ $(<1)$ the
detection probabilities of momenta
$\p_1,\dots,\p_p$ are correlated (anticorrelated). 
The $p$-th order momentum distribution 
$\rho^{(p)}(\p_1,\dots,\p_p\vert \p_1,\dots,\p_p;t)$  
depends on the entire $p$-th order RDM 
$\rho^{(p)}(\r_1,\dots,\r_p\vert \r'_1,\dots,\r'_p;t)$, see Appendix \ref{A}. 
Thus, $g^{(p)}(\p_1,\dots,\p_p,\p_1,\dots,\p_p;t)$ provides 
information about the coherence of $\vert\Psi(t)\rangle$ which is 
not contained in $g^{(p)}(\r_1,\dots,\r_p,\r_1,\dots,\r_p;t)$.

In Young double slit experiments using non-interacting bosons  
$\vert g^{(1)}(\r'_1,\r_1;t_0)\vert=1$  ensures the maximal 
fringe  visibility of an interference pattern. Here, $t_0$ is the time of release from the slits.
See, e.g. \cite{BHE00}, for an experiment using Bose-Einstein condensates. 
However, if interactions during the expansion behind the slit are not negligible, there is no simple 
relation between the fringe visibility and the 
wave function $\Psi(\r_1,\dots,\r_N,t_0)$ at 
the time of release from the slits. In other words,
the interaction between the particles can 
modify the observed interference pattern \cite{CSB06,XSM06,CSB07}.

In order to determine the degree of coherence of a given system, 
it is necessary to quantify how well Eq.~(\ref{coh}) is satisfied.
A visualization of the degree of coherence is
highly desirable, as it helps to understand the
coherence limiting factors in an intuitive manner.
Already for one-dimensional systems, $D=1$, the normalized first-order 
correlation function at time $t$, $g^{(1)}(\r'_1,\r_1;t)$, is a 
complex function of two variables and 
cannot be visualized in a single plot.
It is, therefore, necessary to consider quantities that sample
parts of $g^{(p)}(\r'_1,\dots,\r'_p,\r_1,\dots,\r_p;t)$.
In one-dimensional systems $\vert g^{(1)}(\r_1',\r_1;t)\vert^2$ 
can be represented as a two-dimensional plot and gives a measure 
for the degree of first-order coherence. 
Similarly,  $g^{(2)}(\r_1,\r_2,\r_1,\r_2;t)$ and 
$g^{(2)}(\p_1,\p_2,\p_1,\p_2;t)$ are real and 
can be represented as two-dimensional plots, if $D=1$.
In Secs.~\ref{firstOrderCorrelations} and \ref{secondOrderCorrelations}
we will visualize the degree of first- and second-order coherence 
of a one-dimensional system, defined in Sec.~\ref{model}, by means of
$\vert g^{(1)}(\r_1',\r_1;t)\vert^2$, 
$g^{(2)}(\r_1,\r_2,\r_1,\r_2;t)$ and $g^{(2)}(\p_1,\p_2,\p_1,\p_2;t)$.

\section{NUMERICAL METHODS}\label{sectheory}
The main goal of this work is to investigate exactly the behavior of
first- and second-order correlation functions in interacting many-body systems. 
This requires the computation of the exact many-body wave function
which is generally a difficult problem to solve. 
In some cases, when the general form of the wave function is known
{\em a priori}, an exact solution can be obtained, 
either by solving transcendental equations or by exploiting  
mapping theorems, see e.g. \cite{Gir60,LiL63,McG64,GiW00,YuG05,Gir06,SSA05,PeB07,SDD07,BPG07}.
However, in general it is necessary to solve the full
many-body Schr\"odinger equation numerically in an efficient way.
In Sec.~\ref{mctdhbexp} we give a brief account of the numerical 
method MCTDHB to solve the interacting many-boson problem.
In order to find out to which extent mean-field methods are applicable 
to bosonic, interacting many-body systems, we compare our many-body 
results with those based on mean-field approaches, namely the 
commonly used Gross-Pitaevskii  
mean-field \cite{Gro61,Pit61,PeS02} and the {\em best mean-field} (BMF), 
which we describe briefly in Sec.~\ref{bmfexp}.
It is beyond the scope of this work to explain
either MCTDHB or BMF in detail and we refer the reader to 
Refs. \cite{SAC07,ASC07c,ASC07b} and Refs. \cite{CeS03,SCM04,ASC05a,ASC05b,AlC05}  
for more detailed explanations of MCTDHB and BMF, respectively.

\subsection{The many-body wave function}\label{mctdhbexp}
The exact wave function of an interacting $N$-boson problem can always be expanded
in any complete set of permanents of $N$ particles. Each of the permanents
is constructed from a complete set of single-particle 
functions which are commonly referred to as {\em orbitals}.
Practical computations can {\em never} be carried out in 
complete basis sets and therefore, it is crucial to cut the basis set 
carefully. 

Our starting point is the Schr\"odinger picture 
field operator $\hat{\mathbf \Psi}(\r)$
satisfying the usual bosonic commutation relations Eqs.~(\ref{com_rel}).
It is convenient to expand the field operator in a complete set of
{\em time-dependent} orthonormal orbitals,
\beq\label{annihilation_def}
 \hat{\mathbf \Psi}(\r)=\sum_k \hat c_k(t)\phi_k(\r,t),
\eeq
where the time-dependent annihilation and creation operators 
obey the usual commutation relations 
$\hat c_k(t) \hat c^\dag_j(t) - \hat c^\dag_j(t) \hat c_k(t) = \delta_{kj}$
for bosons at any time. Note that it is not necessary to specify the shape 
of the orbitals at this point. 

The many-body Hamiltonian (\ref{genHam}) is standardly written in second quantized form as 
\beq\label{MB_hamil}
 \hat H = \sum_{k,q} \hat c_k^\dag \hat c_q h_{kq} + 
\frac{1}{2}\sum_{k,s,l,q} \hat c_k^\dag \hat c_s^\dag \hat c_l \hat c_q W_{ksql}, 
\eeq
where the matrix elements of the one-body Hamiltonian $h(\r)$ 
and two-body interaction potential $W(\r-\r')$  
are given by 
\beqn\label{matrix_elements}
 h_{kq}(t) &=& \int \phi_k^\ast(\r,t)  h(\r) \phi_q(\r,t) d\r, \nonumber \\
 W_{ksql}(t) &=& \int \!\! \int \phi_k^\ast(\r,t) \phi_s^\ast(\r',t) W(\r-\r')  
 \phi_q(\r,t) \phi_l(\r',t) d\r d\r'.
\eeqn

The {\it ansatz} for the many-body wave function $\vert\Psi(t)\rangle$ in MCTDHB 
is taken as a linear combination of time-dependent permanents
\beqn\label{MCTDHI_Psi}
& & \!\!\! \left|\Psi(t)\right> = 
 \sum_{\vec{n}}C_{\vec{n}}(t)\left|n_1,n_2,\ldots,n_M;t\right>, \nonumber \\
& & \!\!\! \ \ \left|n_1,n_2,\ldots,n_M;t\right> = \frac{1}{\sqrt{n_1!n_2!\cdots n_M!}} 
\left(\hat c_1^\dag(t)\right)^{n_1}\left(\hat c_2^\dag(t)\right)^{n_2}
\cdots\left(\hat c_M^\dag(t)\right)^{n_M}\left|vac\right>, 
\eeqn
where $\left|n_1,n_2,\ldots,n_M;t\right>$ is assembled from the 
time-dependent orbitals above.
The summation in (\ref{MCTDHI_Psi}) runs over all $\binom{N+M-1}{N}$ permanents
generated by distributing $N$ bosons over $M$ orbitals.
We collect the occupations in the vector $\vec{n}=(n_1,n_2,\ldots,n_M)$,
where $n_1+n_2+\ldots+n_M=N$.   
Of course, if $M$ goes to infinity then the {\it ansatz} (\ref{MCTDHI_Psi}) 
for the wave function becomes exact since the set of 
permanents $\left|n_1,n_2,\ldots,n_M;t\right>$ 
spans the complete $N$-particle Hilbert space.
In practical computations we have to restrict the number $M$
of orbitals from which the permanents $\left|n_1,n_2,\ldots,n_M;t\right>$
are assembled. 
By substituting the many-body {\it ansatz} (\ref{MCTDHI_Psi})
into the action functional of the time-dependent Schr\"odinger equation,
it is possible to derive a coupled set of equations of motion containing
the coefficients $C_{\vec{n}}(t)$ and the set of 
time-dependent orbitals $\phi_k(\r,t)$. The equations
are obtained by requiring the stationarity of the action functional with respect 
to variations of the coefficients  $C_{\vec{n}}(t)$ {\em and} the set of time-dependent 
orbitals $\phi_k(\r,t)$. These coupled equations have to be solved simultaneously, 
leading to an efficient wave package propagation method for bosons \cite{SAC07,ASC07c,ASC07b}.
At first sight it might seem to be an unnecessary complication 
to allow the orbitals $\phi_k(\r,t)$ to depend on time. However,
this additional degree of freedom allows both,
the basis of one-particle functions $\phi_k(\r,t)$ and the coefficients 
$C_{\vec{n}}(t)$ to be variationally optimal at any time.
Note that this is fundamentally different to a
multi-mode {\em ansatz} with fixed orbitals in which the quality 
of the chosen basis set may deteriorate as the system evolves in time.

In order to investigate stationary properties of Bose-Einstein condensates, 
we use a many-body relaxation method. 
By propagating a given initial guess in imaginary time with MCTDHB, 
the system relaxes to the ground state, which 
allows us to treat stationary systems as well. 
A necessary requirement for this procedure to work 
is that the initial guess 
has non-zero overlap with the ground state.
The variational principle  ensures that the set 
of orbitals $\phi_k$ is variationally optimal, 
in the sense that the lowest ground state energy within the 
Hilbert subspace of $N$ bosons distributed over $M$ orbitals is obtained, see in 
this respect Ref. \cite{SAC06}.
We will not go any further into the details of MCTDHB and refer 
the reader to the literature \cite{SAC07,ASC07c,ASC07b}.

\subsection{Best mean-field}\label{bmfexp}
The exact many-body wave function of a bosonic system of $N$ particles
can always be expanded in an infinite weighted sum over
any complete set of permanents of $N$ particles.
In mean-field theory the exact many-body wave function is approximated 
by a single permanent. This single permanent is built from 
a number $M\le N$ of orthogonal orbitals in 
which the $N$ bosons reside.
In the field of Bose-Einstein condensates one
particular mean-field, the Gross-Pitaevskii (GP) mean-field,
has proven to be very successful.
In analogy to non-interacting BECs,
in GP theory it is assumed that the many-body wave function
is given by a single permanent in which  all
particles reside in {\em one} orbital, i.e. $M=1$. 
A minimization of the energy functional with the GP {\em ansatz}
wave function leads to the famous Gross-Pitaevskii equation 
\cite{Gro61,Pit61,PeS02}. The solution of the GP equation 
yields the single orbital 
from which the GP mean-field 
permanent is constructed.

However, it has been shown \cite{CeS03,SCM04,ASC05a,ASC05b,AlC05} 
that the GP mean-field is not
always the energetically-lowest mean-field solution.
The assumption that {\em all} particles 
occupy the same orbital is too restrictive. 
Especially in multi-well trapping geometries
the energetically-lowest mean-field solution 
can be fragmented \cite{CeS03,SCM04,ASC05a,ASC05b,AlC05}, 
see also Sec.~\ref{secbasic}.

In order to obtain the energetically-lowest  
mean-field solution, it is necessary that
the ansatz for the  wave function 
is of the most general mean-field form.
Due to the variational principle, 
the minimization of the respective energy 
functional with respect to 
all parameters of the ansatz wave function
will then give the {\em best } solution within mean-field theory.
It is therefore legitimate to call this mean-field solution the best mean field (BMF).
A procedure to obtain the best mean-field (BMF) solution numerically
has been developed recently \cite{CeS03,SCM04,ASC05a,ASC05b,AlC05}. 

In the best mean-field approach the ansatz for the wave function 
$\vert\Psi\rangle$
is taken as a single 
permanent of $N$ bosons distributed over $M$ time-independent 
orthonormal orbitals $\phi_k(\r)$:
\beqn\label{BMF_Psi}
& & \!\!\! \left|\Psi\right> = 
 \left|n_1,n_2,\ldots,n_M\right>. 
\eeqn
Using this ansatz for the wave function, the energy functional is minimized 
by a variation over the number of orbitals $M$, the occupation numbers $n_i$
and the orbitals $\phi_k(\r)$ themselves \cite{CeS03,ASC05b}.
The variation leads to a set of coupled non-linear equations that have to be solved
to obtain the BMF solution. Thereby, the energetically most favourable 
permanent is selected
to approximate the 
true many-body wave function.
The GP mean-field is contained in the BMF ansatz as can be seen 
by restricting the number of orbitals
to $M=1$.

\section{A MODEL AND ITS PHYSICS}\label{model}
In order to examine correlation functions of Bose-condensed systems, we now turn 
to a specific example. For simplicity we work in one dimension, $D=1$, and 
henceforth we substitute $\r=\x$ and $\p=\k$. We will study the correlation functions
of $N=1000$ repulsively interacting bosons in a double-well 
trap at various barrier heights. 
The dynamics of a similar system has been investigated recently 
in the context of a dynamically raised barrier \cite{SAC07}.  
In order to isolate physical effects that are 
due to the trapping geometry and not 
due to dynamical parameters such as the rate at 
which the barrier is raised,  etc., we restrict our discussion 
to the ground state at different barrier heights.
The restriction to a stationary state allows us to 
omit the time argument in all physical quantities from now on. 
Double-well systems have 
the interesting property that depending on the height 
of the barrier and/or the interaction strength, 
the ground state undergoes a transition from 
a single to a fragmented 
condensate \cite{SpS99,SCM04,ASC05b,SAC06}.
We shall show how this transition from 
a condensed state to a fragmented condensate manifests itself
in the correlation functions.

We work with a dimensionless Hamiltonian of the form 
\begin{equation}\label{hamil}
H=\sum_{i=1}^N\left(-\frac{1}{2}\frac{\partial^2}{{\partial x_i}^2}+V(x_i)\right) + \lambda_0\sum_{i<j}^N\delta(x_i-x_j),
\end{equation}
to solve the stationary Schr\"odinger equation $H\Psi=E\Psi$.  
All quantities in Eq.~(\ref{hamil}) are dimensionless
and the connection to a dimensional Hamiltonian 
\begin{equation}\label{hamiltonian}
\tilde{H}=\sum_{i=1}^N\left(-\frac{\hbar^2}{2m}\frac{\partial^2}{{\partial \tilde{x}_i}^2}+\tilde{V}(\tilde{x}_i)\right) 
	+ \tilde{\lambda}_0\sum_{i<j}^N\delta(\tilde{x}_i-\tilde{x}_j)
\end{equation}
is made by the relations $x_i=\tilde{x_i}/L$, where $L$ is a length scale, 
$V(\x_i)=\frac{mL^2}{\hbar^2}\tilde{V}(Lx_i)$, $\lambda_0=\frac{mL}{\hbar^2}\tilde{\lambda}_0$,
$\Psi(\x_1,\dots,\x_N)=\tilde{\Psi}(\x_1 L,\dots,\x_N L)$ and $E=\tilde{E}\frac{mL^2}{\hbar^2}$.
As an external potential we choose a harmonic trap with an additional central barrier of variable height 
$V(\x_i)=\frac{1}{2}\x_i^2 + A e^{(\frac{-\x_i^2}{2\sigma^2})}$, where $A$ is 
the height of the potential barrier 
and $\sigma=2$ a fixed width.
For the strength of the dimensionless interparticle interaction we choose $\lambda_0=0.01$. 
In the computations using MCTDHB we restrict the number of orbitals to $M=2$, 
yielding a total of $\binom{N+M-1}{N}=1001$ permanents. 

\subsection{Condensed state}\label{condensedState}
\begin{figure}[ht]
    \centering
    \includegraphics[height=140mm,angle=0]{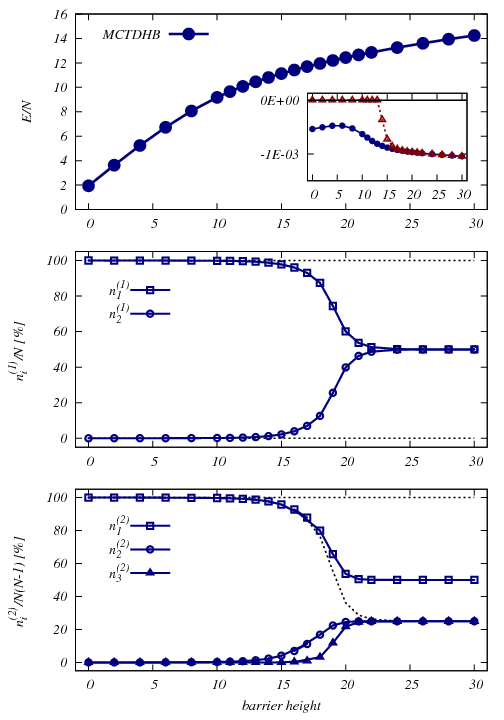}
    \caption{	(color online). Energy per particle, natural orbital and natural geminal occupations
		of the ground state of $N=1000$ bosons at $\lambda_0=0.01$ 
		in a harmonic trap with a central barrier. Shown is the dependence on the barrier height.
		Top: energy per particle $E/N$ of the many-body solution. Inset: 
		energy difference per particle  between the best
		mean-field and the GP solution, $(E_{BMF}-E_{GP})/N$ (triangles), 
		and between the many-body and GP solution, $(E_{MCTDHB}-E_{GP})/N$ (circles).
		Middle: the eigenvalues $n^{(1)}_1$ and $n^{(1)}_2$ 
		of the first-order RDM $\rho^{(1)}(\x_1 \vert\x_1')$. The ground state fragments
		with increasing barrier height.
		Bottom: the eigenvalues 
		$n^{(2)}_1$,$n^{(2)}_2$ and $n^{(2)}_3$ of the second-order RDM
		$\rho^{(2)}(\x_1,\x_2\vert\x_1',\x_2')$.
		The dashed lines in the middle and bottom panel indicate
		upper and lower bounds on the largest eigenvalue 
		of the first- and second-order RDMs. See text for details.
		The quantities shown are dimensionless.}\label{fig1}
\end{figure}

We begin with a discussion of the ground state energy as a function of the barrier height.
The ground state energy per particle of 
the many-body solution, $E_{MCTDHB}/N$ (blue), 
is shown in Fig.~\ref{fig1} (top).
$E_{MCTDHB}/N$ increases with
the height of the central barrier.
The energy differences per particle
of the many-body and the BMF solution with respect
to the GP solution,
$(E_{MCTDHB}-E_{GP})/N$ (blue) and $(E_{BMF}-E_{GP})/N$ (red),
are shown in the inset of Fig.~\ref{fig1} (top).
The energy difference $(E_{MCTDHB}-E_{GP})/N$ is negative, 
because the interacting system can lower its energy 
by depleting the condensate.
At low barrier heights the GP mean-field 
is the best mean-field and thus
$(E_{BMF}-E_{GP})/N=0$.
A comparison of the energy scales of Fig.~\ref{fig1} and its inset reveals
that the energy of the many-body solution, the BMF solution and the GP
solution are very close at all barrier heights.
The nature of the many-body ground state at different barrier
heights varies nevertheless very strongly, 
as we shall show below.

Fig.~\ref{fig1} (middle) shows the occupations $n^{(1)}_1/N$ and $n^{(1)}_2/N$ 
of the first and second natural orbitals of the many-body wave function 
as a function of the barrier height, computed with MCTDHB.
The largest eigenvalue of the first-order RDM, $n^{(1)}_1$, 
is only restricted by Eq.~(\ref{upperbound})
and can therefore take on any value between $0$ and $N$. 
The dashed lines indicate these upper and lower bounds on 
$n^{(1)}_1$. 
At low barrier heights only one natural orbital is significantly occupied.
Therefore, we refer to the parameter 
range $0\le A\le13$ as the condensed regime, in accordance with the definition
of Penrose and Onsager \cite{PeO56}.
The occupation of the second natural orbital is due to the two-body interaction 
between the particles.
However, it remains below $1\%$ for all values of the barrier height $A\le13$ and 
is even below $1\promille$ at $A=0$.

Since $n_1^{(1)}\approx N$ in the condensed regime, 
the upper and the lower bounds, 
Eqs.~(\ref{upperbound}) and  (\ref{lowerbound}), 
on  the largest eigenvalue 
of the second-order RDM, $n^{(2)}_1$,  
are almost identical. 
Therefore, $n^{(2)}_1$ is constrained to take on 
a value very close to $N(N-1)$. 
Consequently, there can be only one 
significantly occupied natural geminal.
This is confirmed in Fig.~\ref{fig1} (bottom), 
where the natural geminal occupations
are shown as a function of the barrier height. 
For the purpose of describing first- and 
second-order correlations it is therefore 
legitimate to approximate the many-body 
wave function in this regime by a single permanent
$\vert N,0\rangle$ in which all $N$ bosons occupy 
the first natural orbital $\alpha_1^{(1)}(\x_1)$. 

The first column of Fig.~\ref{fig2} shows the first (red) 
and the second (blue)  natural orbitals of the many-body solution
at barrier heights $A=0,13,19,24$, from top to bottom.
The first and the second natural orbitals are
symmetric and antisymmetric about the origin, respectively.
At $A=0$ the first natural orbital, $\alpha_1^{(1)}(\x_1)$, 
takes on the shape of a broadened Gaussian, 
reflecting the repulsive interaction between the particles.
The second natural orbital, $\alpha_2^{(1)}(\x_1)$, 
has a higher kinetic energy than the first one due to the node 
at the center of the trap. 
Additionally, the second natural orbital forces the particles to occupy
regions of the trap where the trapping
potential is higher. There is an energy gap
between the one-particle energies of the first 
and second natural orbital. The occupation of the second 
natural orbital is therefore 
very small in the purely harmonic trap 
at the chosen interaction strength.

As the barrier height is varied from $A=0$ up 
to $A=13$, the natural orbitals deform
to fit the new shape of the external potential.
The central peak of the first natural orbital splits into two  
maxima which become localized at positions $\x_1=\pm d/2$, where $d$
is the distance between the wells of the 
external potential. 

At the center of the trap, 
where the barrier is raised, 
the first natural orbital develops a local minimum 
in order to minimize the potential energy.
The second natural orbital on the other hand
has a node at the center of 
the trap at any barrier height.
Its maximum and  minimum are 
localized at the  minima of the external potential.
As the barrier is raised, 
the energy gap between the first two natural orbitals
decreases. However, the increase of the depletion of the condensate 
from $n_2^{(1)}/N< 1\promille$ at $A=0$ to $n_2^{(1)}/N\approx 1\%$ at $A=13$
cannot be explained in this single particle picture.
On a single particle level 
the ground state would be fully condensed at any
finite barrier height, i.e. $n_2^{(1)}=0$.
The reason for the observed increase in 
the depletion lies in the fact 
that for repulsively interacting many-boson systems in multi-well
setups it becomes energetically more favourable 
to fragment as the barrier between 
the wells is raised \cite{SpS99,SCM04,ASC05a,ASC05b,SAC06}.
The increase in energy which results from the occupation of orbitals
with a higher one-particle energy can be outweighed by a 
decrease in interaction energy. 
This effect becomes dominant at barrier
heights above $A=13$, see Secs.~\ref{manyBodyState} and \ref{fragmentedState}.
 
The second to fourth column of Fig.~\ref{fig2} show (from left to right) 
the first three natural geminals 
$\alpha^{(2)}_i(\x_1,\x_2)$ at the same barrier
heights as above. By expanding the natural geminals 
in the one-particle basis of natural orbitals
one finds that the natural geminals in the condensed regime 
are approximately given by symmetrized 
products of the natural orbitals:
\begin{eqnarray}\label{condensedgems}
\vert\alpha^{(2)}_1\rangle=\vert 2,0\rangle,\,\,\,\,\,\,\,
\vert\alpha^{(2)}_2\rangle=\vert 1,1\rangle,\,\,\,\,\,\,\,
\vert\alpha^{(2)}_3\rangle=\vert 0,2\rangle,
\end{eqnarray}
where $\vert m_1,m_2\rangle$ denotes a 
state with $m_1$ particles in the first and $m_2$ 
particles in the second natural orbital.
Only  the first natural geminal, $\alpha_1^{(2)}(\x_1,\x_2)$, 
is significantly occupied in the condensed regime.
Due to the two-body interaction between the particles there is a small 
occupation of the second and third natural geminal.
However, at low barrier heights their occupation is largely suppressed, 
due to the gap between the single particle energies 
of the first and second natural orbital.
Since the geminals $\alpha_2^{(2)}(x_1,x_2)$ 
and $\alpha_3^{(2)}(x_1,x_2)$ contain 
the second natural orbital in their expansion, see Eq.~(\ref{condensedgems}), 
their occupation increases the 
total energy at low barrier heights.

In the equation for the energy 
expectation value,  Eq.~(\ref{gemEnergy}), 
the only substantially contributing natural geminal 
is $\alpha_1^{(2)}(x_1,x_2)$.
The shape of $\alpha_1^{(2)}(x_1,x_2)$ is particularly interesting.
It has four maxima of similar height, located at positions $\x_1=\x_2=\pm d/2$ and
$\x_1=-\x_2=\pm d/2$, see the second panel in the second row of Fig.~\ref{fig2}.
Since $\alpha_1^{(2)}(x_1,x_2)$ has peaks on the diagonal 
at $\x_1=\x_2=\pm d/2$, it contributes to both, 
the one-particle part and the interaction part of the energy. 

In contrast to the first natural geminal, $\alpha_2^{(2)}(x_1,x_2)$ 
and  $\alpha_3^{(2)}(x_1,x_2)$ both exhibit node lines 
going through the region  where the central barrier is raised. 
As the energy gap between the single particle energies 
of the natural orbitals $\alpha_1^{(1)}(x_1,x_2)$ and $\alpha_2^{(1)}(x_1,x_2)$ 
decreases, so does the energy gap between the natural geminals.
Similar to the discussion of the natural orbital occupations above, 
this argument in terms of an energy gap does not explain the increase of the
occupation of the second and third natural geminal 
when the barrier is raised. Without interactions
the occupation numbers of all 
but the first natural geminal would be exactly zero.

We shall demonstrate in Sec.~\ref{fragmentedState} that fragmented 
states allow the occupation
of geminals that contribute very little 
to the interaction energy
as opposed to condensed states. 
Thereby, the system can lower its  energy, 
once the barrier is high enough.

\subsection{From condensation to fragmentation}\label{manyBodyState}
At barrier heights  $13<A<24$, one finds 
that the occupation of the second natural orbital $n_2^{(1)}/N$
increases continuously from below $1\%$ to almost $50\%$.
The condensate fragments in this regime according to the definition of 
fragmented condensates \cite{NoS82}.
In this regime many permanents contribute to the wave function and, 
therefore, we refer to the range
of barrier heights $13<A< 24$ as the many-body regime.  
Along with the natural orbital occupations the natural geminal occupations change as well. 
Three natural geminals become occupied with increasing barrier height, see Fig.~\ref{fig1}.

In the many-body regime, 
the upper and lower bounds, Eqs.~(\ref{upperbound}) and (\ref{lowerbound}), on the 
largest eigenvalue of the second-order RDM, $n^{(2)}_1$, no longer restrict 
$n^{(2)}_1$ to a narrow region.
In fact, $n^{(2)}_1$ takes on a value somewhere in between these bounds. 
 
This onset of fragmentation manifests itself also 
in the BMF solution which jumps from a GP type 
permanent $\vert N,0\rangle$ in the condensed regime 
to a fully fragmented solution of the 
form $\vert N/2,N/2\rangle$.
Note that already at barrier heights $A\ge14$ this fragmented solution is lower in 
energy than a GP type permanent. At this barrier height
the many-body solution is only slightly depleted, see Fig.~\ref{fig1}. 

If we compare the natural orbitals in Fig.~\ref{fig2} at barrier height $A=19$ with those at $A=13$,
we note that they look very similar, apart from the fact that the peaks are
slightly farther apart, and the first natural orbital is closer to zero at the center of the trap.
The energies of the first two natural orbitals are almost degenerate and 
the total energy is minimized by the occupation of both natural orbitals. 
Without interactions the system would remain in a condensed state, 
since the single particle energies of the first and the second natural orbitals 
remain separated at any finite barrier height. Note that in the 
absence of interactions the natural orbitals are the eigenfunctions of $\hat{h}$.
However, as we noted in Sec.~\ref{condensedState}, a system of repulsively
interacting bosons in multi-well traps can lower its energy by occupying several natural  
orbitals, once the barrier is high enough \cite{SCM04,ASC05a,ASC05b,SAC06,AlC05}.  
This is precisely the reason for 
the  observed onset of fragmentation.

In the many-body regime the natural geminals are no longer symmetrized products
of the natural orbitals. 
If we compare $\alpha^{(2)}_1(\x_1,\x_2)$ in Fig.~\ref{fig2} at barrier heights $A=13$ and $A=19$, 
we see that the peaks on the diagonal at $\x_1=\x_2=\pm d/2$
decrease, whilst the peaks on the off-diagonal at $\x_1=-\x_2=\pm d/2$
increase when the barrier is raised. 
The opposite is true for the third natural geminal $\alpha^{(2)}_3(\x_1,\x_2)$: the 
off-diagonal maxima at $\x_1=-\x_2=\pm d/2$ have decreased, 
whilst the diagonal minima at $\x_1=\x_2=d/2$ are now more negative.
On the other hand, the second natural geminal, $\alpha^{(2)}_2(\x_1,\x_2)$, is still well approximated by
a symmetrized product of the first and second natural orbital.
The  behavior of the natural geminals is    
qualitatively different from that displayed by the natural orbitals.
In contrast to the natural orbitals, the natural geminals {\em do} change their shape 
during the fragmentation transition. They only obtain their final forms, 
when the fragmentation transition is completed, see Fig.~\ref{fig2} 
and Sec.~\ref{fragmentedState}. 
It is possible
to understand qualitatively which of the two unoccupied 
natural geminals becomes occupied first. 
In the condensed regime $\alpha_2^{(2)}(x_1,x_2)$ is well approximated
by a state $\vert 1,1\rangle$ in which one boson occupies the first natural orbital
and another one the second natural orbital, see Eq.~(\ref{condensedgems}).
Meanwhile, $\alpha_3^{(2)}(x_1,x_2)$ is well approximated by a state  
$\vert 0,2\rangle$, in which two bosons reside in the second natural orbital.
Since the single particle energies of the first and second natural orbitals are separated,
the occupation of the third natural geminal costs more energy than that of the second.
Hence the second natural geminal becomes occupied before the third, see Fig.~\ref{fig1} (bottom).

\subsection{Fully fragmented state}\label{fragmentedState}
When the central barrier is raised to values $A\ge24$, 
the two parts of the condensate become truly independent. 
The natural orbital occupations approach $n_1^{(1)}=n^{(1)}_2=N/2$, which reflects
the fact that the energies associated with the 
first and second natural orbitals degenerate at infinite barrier heights.
The many body wave function can then be adequately approximated 
by a single permanent of the form $\vert N/2,N/2\rangle$, 
i.e. with equal numbers of particles 
in the first and the second natural orbitals.
Therefore, we refer to barrier heights $A\ge24$ as the fully fragmented regime.
The additional  energy, necessary for the occupation of the second natural orbital, is outweighed 
by a lower interaction energy.
Note that this final form of the 
wave function is anticipated by the BMF 
solution at barrier heights $A\ge 14$. 

The natural geminal occupations approach 
\begin{eqnarray}\label{fragGemOcc}
n_1^{(2)}= N(N/2),\,\,\,\,\,\,\,n_2^{(2)} = n_3^{(2)}= N/2(N/2-1)
\end{eqnarray}
in the fully fragmented regime. 
These are the values that follow 
from the BMF solution. 

\begin{figure}
    \centering
    \includegraphics[width=150mm,angle=0]{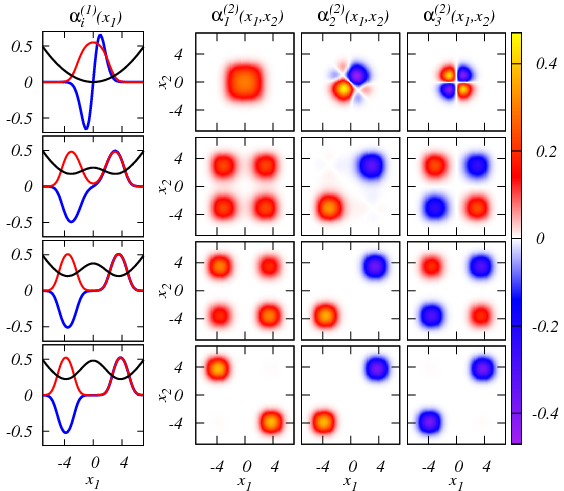}
    	\caption{(color online). Natural orbitals and geminals at different barrier heights. 
		First column: the natural orbitals $\alpha_1^{(1)}(\x_1)$ (red line) 
	    	and $\alpha^{(1)}_2(\x_1)$ (blue line) of the many-body solution at different 
		barrier heights $A=0,13,19,24$, from top to bottom. The  
		trapping potential is shown as a black line in the first column.	
		The state of the system changes from condensed to fragmented between $A=13$ and $A=24$. 
		Second to fourth columns:
		natural geminals $\alpha^{(2)}_1(\x_1,\x_2)$, $\alpha^{(2)}_2(\x_1,\x_2)$ and $\alpha^{(2)}_3(\x_1,\x_2)$ 
		from left to right at the same barrier heights as above. While the natural orbitals
		remain qualitatively unchanged during the fragmentation transition, 
		the natural geminals take on their final shapes only when 
		the system becomes fully fragmented. 
		At low barrier heights only one natural geminal is occupied.
		At high barriers three natural geminals are occupied, see Fig.~\ref{fig1}. 
		The total energy is 
		minimized by the occupation of a natural geminal that
		contributes practically nothing to the interaction energy. 
		See text for more details. The quantities shown are dimensionless.}
    \label{fig2}
\end{figure}

It is only at barrier heights $A\ge24$ that the natural geminals 
take on their final shapes,
compare the third and fourth rows of Fig.~\ref{fig2}.
If we expand the natural geminals in the basis 
of natural orbitals at these barrier heights, we find that 
\begin{eqnarray}\label{fragmentedgems}
\vert\alpha^{(2)}_1\rangle=\frac{1}{\sqrt{2}}\left(\vert 2,0\rangle-\vert 0,2 \rangle\right),\,\,\,\,\,\,\,
\vert\alpha^{(2)}_2\rangle=\vert 1,1\rangle,\,\,\,\,\,\,\,
\vert\alpha^{(2)}_3\rangle=\frac{1}{\sqrt{2}}\left(\vert 2,0\rangle+\vert 0,2 \rangle\right)
\end{eqnarray}
holds to a very good approximation. 
The first and third natural geminals have equal contributions 
coming from the first and the second natural orbitals. The question arises, 
why their occupations are different, about $50\%$ and $25\%$, respectively.
Subtracting the permanents 
$\vert 2,0\rangle$ and $\vert 0,2 \rangle$ from one another yields
a geminal which is localized 
on the {\em off-diagonal}, see $\alpha_1^{(2)}(x_1,x_2)$ in Fig.~\ref{fig2} at $A=24$.
Adding the permanents $\vert 2,0\rangle$ 
and $\vert 0,2 \rangle$ 
yields a geminal which is localized on the 
{\em diagonal}, see $\alpha_3^{(2)}(x_1,x_2)$ in Fig.~\ref{fig2} at $A=24$.
It is easy to see from the shape of the 
natural geminals in the fourth row of Fig.~\ref{fig2}
that the integrals over the one-body part in Eq.~(\ref{gemEnergy}) 
are approximately the same for each of the natural geminals.
Given the occupations in Eq.~(\ref{fragGemOcc}), the first 
natural geminal contributes about one half of the 
one-body energy, whereas the 
second and the third natural geminal contribute about a fourth each.
The situation is different for the two-body part of the Hamiltonian.
Since $\alpha_2^{(2)}(x_1,x_2)$ and $\alpha_3^{(2)}(x_1,x_2)$ are localized on the diagonal,
they {\em do} contribute to the interaction energy. 
In contrast, $\alpha_1^{(2)}(x_1,x_2)$ is almost zero at coordinate values 
$\x_1\approx\x_2$ and, due to the contact interaction in Eq.~(\ref{hamiltonian}), 
it practically does not contribute to the interaction energy. 
At high barriers a fragmented state allows 
the system to lower its energy
through the occupation of a natural geminal
which is localized on the off-diagonal.

We would like to make a remark on the validity of the present 
MCTDHB computation for high barriers.
For high barriers the whole system can 
be considered as composed of two separate condensates.
To describe the depletion of each condensate 
it would be necessary to employ $M=4$ orbitals.
We use only $M=2$ orbitals in the many-body computation 
and cannot describe 
this depletion. We justify the use of $M=2$ 
orbitals by noting that at $A=0$
the system is almost fully condensed, and 
the depletion can be safely neglected, 
see Fig.~\ref{fig1}. 
Therefore, we assume that the
depletion of each of the two condensates 
can be neglected, when the barrier is very high.
This claim is supported by a computation that we carried out in the 
harmonic trap at the same interaction strength $\lambda_0=0.01$ 
for $500$ particles. 
The depletion was found to 
be even less than for $N=1000$ particles.

\section{FIRST ORDER CORRELATIONS}\label{firstOrderCorrelations}
\subsection{General analytical considerations}\label{generalResults1}
We now describe the first-order correlations in an analytical 
mean-field model for the two limiting cases
of a condensed and a fully fragmented system.
In these cases mean-field theory has been shown to be well applicable, 
see Sec.~\ref{model}.
For our purposes the exact shape of the natural orbitals 
$\alpha^{(1)}_1(\x_1)$ and $\alpha^{(1)}_2(\x_1)$
is unimportant.
Consider a normalized one-particle function, $\Phi(\x)$,
which is localized at the origin. $\Phi(x)$ may vary in shape, but 
is always assumed to resemble a 
Gaussian. Similarly, we define translated copies 
$\Phi_1(x)=\Phi(x+d/2)$ and $\Phi_2(x)=\Phi(x-d/2)$
of $\Phi(x)$, where the previously defined distance $d$ between
the minima of the potential wells is taken to be large enough
to set products of the form $\Phi_1(x)\Phi_2(x)$ 
to zero.
Since $\Phi$ is localized in some region around the origin, 
$\Phi_1$ is localized in a region $L$ to the left
and $\Phi_2$ in a region $R$ to the right of the origin.

\subsubsection{Condensed state}
In the condensed regime, $0\le A\le13$, only one 
natural orbital, $\alpha^{(1)}_1(\x_1)$, is significantly 
occupied. Therefore, we approximate the 
first-order reduced density operator of the 
system by that of a condensed state $\vert N,0\rangle$ 
\begin{equation}\label{condensedRho1Op}
\hat{\rho}^{(1)}_{\vert N,0\rangle}     = N\vert \alpha_1^{(1)}\rangle\langle \alpha_1^{(1)} \vert. 
\end{equation}
It then follows from Eq.~(\ref{gp}) that 
\begin{equation}\label{g1LocalizedCondensed} 
\vert g^{(1)}_{\vert N,0\rangle}(\x_1',\x_1)\vert^2		=	1.
\end{equation}
At zero barrier height, the first natural orbital is 
a Gaussian, broadened by interactions.  
Therefore, we write $\alpha_1^{(1)}(\x_1)=\Phi(\x_1)$, 
and hence the one-particle density distribution and the one-particle 
momentum distribution are of the form
\begin{eqnarray}
\rho^{(1)}_{\vert N,0\rangle}(\x_1\vert\x_1)     &=& N \vert\Phi(\x_1)\vert^{2}  \label{localizedDensityCondensed}, \\
\rho^{(1)}_{\vert N,0\rangle}(\k_1\vert\k_1)     &=& N \vert\Phi(\k_1)\vert^{2}. \label{localizedMomentumCondensed}
\end{eqnarray}
Since $\Phi(\x_1)$ is a broadened Gaussian, its Fourier transform $\Phi(k_1)$ 
is also close to a Gaussian, but narrower in comparison to a non-interacting system.
The momentum distribution of the repulsively interacting system in the harmonic trap 
is therefore narrower than that of a non-interacting system.

We now turn to the case corresponding to 
$A\approx13$, where the system is still condensed, 
but the first two natural orbitals are spread 
out over the two wells. 
We model the natural orbitals by
\begin{eqnarray}\label{distOrbs}
\alpha_1^{(1)}(\x_1)=\frac{1}{\sqrt{2}}\left[\Phi_1(\x_1)+\Phi_2(\x_1)\right],\,\,\,\,\,\,\,
\alpha_2^{(1)}(\x_1)=\frac{1}{\sqrt{2}}\left[\Phi_1(\x_1)-\Phi_2(\x_1)\right].
\end{eqnarray} 
In this case one obtains  \cite{PiS99}:
\begin{eqnarray}
\rho^{(1)}_{\vert N,0\rangle}(\x_1\vert\x_1)  &=& \frac{N}{2}\vert\Phi_1(\x_1)\vert^2 + \frac{N}{2}\vert\Phi_2(x_1)\vert^2,\label{distributedDensityCondensed}\\
\rho^{(1)}_{\vert N,0\rangle}(\k_1\vert\k_1)   &=& N[1+\cos(k_1d)]\vert\Phi(k_1)\vert^2 \label{distributedMomentumCondensed}
\end{eqnarray}
for the density and the momentum distribution.
We note that the one-particle momentum distribution 
displays an oscillatory pattern in momentum 
space at a period which is determined 
by the separation $d$ of the centers of the two wells.

\subsubsection{Fully fragmented state}
In the true many-body regime, $13<A<24$, 
where many permanents contribute to the wave function,
a mean-field model is bound to fail. However, in the 
fully fragmented regime it is possible 
to consider the whole system as two separate condensates, 
and hence a mean-field description is again applicable.
Therefore, we now turn to the case corresponding to $A\ge 24$, where the system is fully fragmented 
and the many-body state is given by $\vert N/2,N/2\rangle$.
The first-order density operator then reads: 
\begin{equation}
\hat{\rho}^{(1)}_{\vert N/2,N/2\rangle} = 
\frac{N}{2}\vert \alpha_1^{(1)}\rangle\langle\alpha_1^{(1)}\vert +
\frac{N}{2}\vert \alpha_2^{(1)}\rangle\langle \alpha_2^{(1)}\vert.
\end{equation}
Since the natural orbitals remain qualitatively unchanged during the fragmentation transition, 
we approximate $\alpha_1^{(1)}(\x_1)$ and $\alpha_2^{(1)}(\x_1)$ by  
Eqs.~(\ref{distOrbs}) and obtain
for the density and the momentum distribution  \cite{PiS99}:
\begin{eqnarray}
\rho^{(1)}_{\vert N/2,N/2\rangle}(\x_1\vert\x_1)  &=& 
\frac{N}{2}\vert\Phi_1(\x_1)\vert^2 + \frac{N}{2}\vert\Phi_2(\x_1)\vert^2      \label{distributedDensityFragmented},\\
\rho^{(1)}_{\vert N/2,N/2\rangle}(\k_1\vert\k_1)  &=& N \vert\Phi(k_1)\vert^2. \label{distributedMomentumFragmented}
\end{eqnarray}
We note that the one-particle 
momentum distribution of independent 
condensates does not contain an oscillatory
component and is identical  
to the momentum distribution of a single, localized condensate 
of $N$ particles within this model, see Eq.~(\ref{localizedMomentumCondensed}).
For the normalized first-order correlation  
function one finds 
\begin{eqnarray}\label{g1DistributedFragmented}
\vert g^{(1)}_{\vert N/2,N/2\rangle}(\x_1',\x_1)\vert^2	&=&	
				\begin{cases}
					1&\text{if } x_1,x_1' \in L \text{ or } x_1,x_1'\in R,\\
					0&\text{otherwise. }\\
				\end{cases} 
\end{eqnarray}
Whereas the state $\vert N,0\rangle$ is fully 
first-order coherent, the fragmented state $\vert N/2,N/2\rangle$
is only first-order coherent in a restricted and generally disconnected region.
Each of the two  condensates is first-order coherent, 
but the mutual coherence which is present in the condensed regime
is lost.

\subsection{Numerical results}\label{numericalResults1}
We now turn to the discussion of first-order correlations. 
In particular, we are interested 
in effects that are due to 
the true many-body nature of the wave function.
Along with our many-body results we plot the corresponding results of the BMF solution.
From the discussion in Sec.~\ref{model} it is clear that we expect many-body effects
to occur during the fragmentation transition at barrier heights $13<A<24$. In the 
condensed and in the fully fragmented regime we expect that the
many-body results are well approximated by those of the BMF solution.
In these cases we can  
understand the structure 
of the correlation functions
on the basis of the analytical 
mean-field model of Sec.~\ref{generalResults1}.

\begin{figure}
    \centering
    \includegraphics[width=130mm,angle=0]{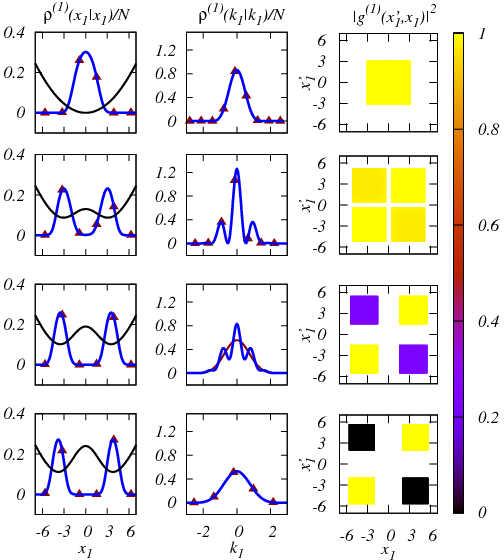}
    	\caption{(color online). Density distribution, momentum distribution and first-order coherence.
		The first two columns show the one-particle density $\rho^{(1)}(\x_1\vert\x_1)/N$ and
		the one-particle momentum distribution $\rho^{(1)}(\k_1\vert \k_1)/N$ of the 
		many-body solution (blue line) and of the BMF solution (red line with triangles), respectively. 
		From top to bottom the height of the 
		central barrier is $A=0,13,19,24$. The BMF result agrees well with the many-body result 
		for a large range of barrier heights. At $A=19$, in the many-body regime,  
		deviations are visible in the momentum distribution. See text for details.
		The third column shows the absolute value squared of the normalized first-order correlation
		function $\vert g^{(1)}(\x_1',\x_1)\vert^2$ at the same barrier heights. 
		An initially coherent condensate splits into two separate condensates which are no longer
		mutually coherent. Only the coherence within each of the wells is preserved. 
		The quantities shown are dimensionless.
		}
    \label{fig3}
\end{figure}

The first column of Fig.~\ref{fig3} shows the one-particle 
density distribution $\rho^{(1)}(\x_1\vert\x_1)$ of the many-body 
solution (blue line) and that of the BMF solution 
(red line with triangles) at the barrier heights 
$A=0,13,19,24$, from top to bottom. 
It is remarkable that the one-particle densities obtained 
from either the many-body wave function or the BMF solution
give results that cannot be distinguished 
from one another at {\em any} barrier height.

In a purely harmonic trap, $A=0$, 
the one-particle density takes on the 
form of an interaction-broadened Gaussian.
At higher barriers, the density splits
into two parts that are localized in each of the wells. 
At $A=13$, the one-particle density 
has developed two separated peaks. 
Note that the system is still in the condensed 
regime at this barrier height and 
must be considered a single condensate, 
despite the spatial separation 
between the two peaks. 

When the central barrier is raised 
further to values $13<A<24$ the condensate
fragments, see Fig.~\ref{fig1}. 
At a barrier height of $A=19$ the 
system is halfway on its way from a 
condensed to a fully fragmented condensate.
Many permanents contribute to the 
many-body wave function and 
one may wonder how this transition manifests 
itself in observable quantities.
However, apart from a small shift of 
the center of the two peaks and a 
reduction  at the center of the trap, 
$\rho^{(1)}(\x_1\vert\x_1)$ remains  largely
unaffected by this transition. 
If the barrier is raised further to $A=24$, 
the fragmentation 
transition is largely completed. 
Also during the transition from a  true many-body state
to a fully fragmented state
there is no visible indication  of this transition
in the one-particle density.

The second column of Fig.~\ref{fig3} 
shows the one-particle 
momentum distribution $\rho^{(1)}(\k_1\vert\k_1)$ 
at the same barrier heights as before. 
At $A=0$, the one-particle 
momentum distribution is given by 
a squeezed Gaussian, in agreement with
Eq.~(\ref{localizedMomentumCondensed}).
At $A=13$ the one-particle
momentum distribution 
has developed an oscillatory pattern,
typical of a single condensate 
spread out over two
wells. The structure of  $\rho^{(1)}(\k_1\vert\k_1)$ is well reproduced by
Eq.~(\ref{distributedMomentumCondensed}) 
of the analytical mean-field model.
Up to this barrier height the BMF solution is 
almost identical to the many-body wave function, and therefore 
the respective momentum distributions 
are indistinguishable, see the two upper panels 
in the second column of Fig.~\ref{fig3}.

When the system enters the many-body regime, $13<A<24$, 
the momentum distribution of the many-body solution deforms  
to a Gaussian-like envelope, modulated by an oscillatory part.
The BMF momentum distribution, on the other hand, 
already takes on the form characteristic
of two separate condensates. 
It agrees with the prediction of
Eq.~(\ref{distributedMomentumFragmented}), which is clearly different 
from the many-body result. This merely reflects the fact that 
the exact wave function is inaccessible 
to mean-field methods in the many-body regime.

When the state becomes fully fragmented at $A=24$,
the many-body momentum distribution and 
the BMF momentum distribution become indistinguishable
again, consistent with an explanation 
in terms of two independent condensates, see
Eq.~(\ref{distributedMomentumFragmented}).
Compared to $\rho^{(1)}(\k_1\vert\k_1)$ at $A=0$,
the momentum distribution is broader at $A=24$,
because the density distribution in each of the 
two wells is  narrower than that in the harmonic trap.

The third column of Fig.~\ref{fig3}
shows the absolute value squared of the normalized 
first-order correlation function 
$\vert g^{(1)}(\x_1',\x_1)\vert^2$ of the many-body solution only. 
Here and in all following graphs of correlation functions
we restrict the plotted region by a simple rule. To avoid analyzing 
correlations in regions of  space 
where the density is essentially zero,
we  plot the respective correlation function only in regions where 
the density is larger than $1\%$ of the maximum 
value of the density in the entire space. 
We apply the same rule also in momentum space.

At zero barrier height $\vert g^{(1)}(\x_1',\x_1)\vert^{2}$ 
is very close to one in the region 
where the density is localized. 
The system is first-order coherent 
to a very good approximation
and the mean-field formula 
Eq.~(\ref{g1LocalizedCondensed}) applies.
As the barrier is raised to $A=13$ 
the coherence between the two peaks, e.g. at $\x_1=-\x_1'$, 
is slightly decreased, while the coherence
within each of the peaks is preserved.
Note that the density at the center of 
the trap is already below $1\%$ of the 
maximal value in this case.
Despite this separation the system remains
largely condensed, but deviations 
from Eq.~(\ref{g1LocalizedCondensed}) 
are visible. If the barrier is raised further to $A=19$, 
the coherence of the system on the 
off-diagonal decreases quickly.
Although the bosons in each well remain coherent among each other, 
the overall system is only partially coherent.
At barrier heights $A\ge24$, 
the coherence between the two
wells is entirely lost. This is also the scenario  that the BMF solution 
anticipates, see Eq.~(\ref{g1DistributedFragmented}).

It is remarkable that not only the density, but also  
the momentum distribution obtained within mean-field theory agree 
so well with the many-body result, when the system is not in the 
true many-body regime.
This would not be the case 
if we had restricted the mean-field
approach to the GP equation, 
as we shall show now.

\begin{figure}
    \centering
    \includegraphics[width=130mm,angle=0]{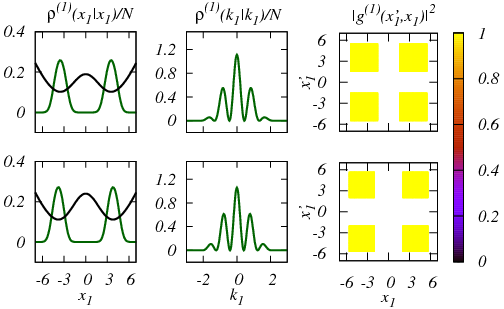}
    	\caption{(color online). Density distribution, momentum distribution and first-order coherence obtained 
		by using the GP equation for high barriers.
		The first two columns show the GP one-particle 
		density $\rho^{(1)}(\x_1\vert\x_1)/N$ (left) and
		the GP one-particle momentum distribution $\rho^{(1)}(\k_1\vert\k_1)/N$ (middle) 
		at barrier heights $A=19$  and $A=24$ (solid green lines).
		In the first column the trapping potential is also shown (solid black line).
		The GP equation models the density well, but fails at
		the computation of the momentum distribution, compare with Fig.~\ref{fig3}.
		The third column shows the absolute value squared of the normalized first-order correlation
		function $\vert g^{(1)}(\x_1',\x_1)\vert^2$ computed with the GP equation 
		at the same barrier heights. The normalized first-order correlation function
		is incorrectly described by the solution of the GP equation.  The quantities shown are dimensionless.
		}
    \label{fig4}
\end{figure}

Up to barrier heights $A=13$ 
the many-body system is condensed, and the BMF solution 
coincides with the GP solution. 
The BMF, and therefore also the GP solution
provide good approximations to the interacting 
many-body system.
Above $A=13$ the results obtained 
with the GP mean-field become qualitatively 
wrong as the barrier is raised.
To illustrate this point, 
we plot the GP results 
corresponding to those of Fig.~\ref{fig3}
at barrier heights $A=19$ (top) and $A=24$ (bottom) 
in Fig.~\ref{fig4}.
A comparison of the respective one-particle 
densities, shown in the first column 
of Fig.~\ref{fig3} and Fig.~\ref{fig4}, 
reveals no visible difference.
The GP mean-field reproduces 
the density distribution at all barrier heights correctly.
However, the GP  solution fails at the description of the 
momentum distribution and the normalized first 
order correlation function, compare the second and third columns
of Figs.~\ref{fig3} and \ref{fig4} at the same barrier heights.
The reason for the failure of the GP mean-field is the
assumption that all bosons
occupy the same orbital. It is 
by construction incapable to describe fragmented condensates.

\section{SECOND ORDER CORRELATIONS}\label{secondOrderCorrelations}
\subsection{General analytical considerations}\label{generalResults2}
In this subsection we extend the 
analytical mean-field model of Sec.~\ref{generalResults1}
to describe second-order correlations.

\subsubsection{Condensed state}
We found in Sec.~\ref{condensedState} that only 
one natural geminal is significantly occupied
in the condensed regime, where the many-body 
state is approximately given by a single
permanent in which all bosons occupy the same 
single particle state.
Therefore, we approximate the second 
order reduced density operator in the condensed regime
by that of the state $\vert N,0\rangle$:
\begin{eqnarray}\label{rho2Op}
\hat{\rho}^{(2)}_{\vert N,0\rangle}     &=& N(N-1)\vert\alpha^{(2)}_1\rangle\langle\alpha^{(2)}_1\vert, 
\end{eqnarray}
where $\alpha^{(2)}_1(\x_1,\x_2)=\alpha_1^{(1)}(\x_1)\alpha_1^{(1)}(\x_2)$ is  
the permanent in which two bosons reside in the first natural orbital $\alpha_1^{(1)}$.
For the condensed state $\vert N,0\rangle$ one finds that
up to corrections of order ${\mathcal O}(1/N)$ 
the state is second-order coherent:
\begin{eqnarray}
g^{(2)}_{\vert N,0\rangle}(\x_1,\x_2,\x_1,\x_2) &=& 1-\frac{1}{N},\label{g2xcondensed}\\
g^{(2)}_{\vert N,0\rangle}(k_1,k_2,k_1,k_2)	&=& 1-\frac{1}{N}.\label{g2kcondensed}
\end{eqnarray}
Thus, there are practically no two-body correlations if $N\gg1$.
At zero barrier height the first natural orbital takes 
on the shape of a broadened Gaussian,
$\alpha_1^{(1)}(\x_1)=\Phi(\x_1)$, where $\Phi(\x)$ is
defined in Sec.~\ref{generalResults1}. 
The first natural geminal then reads
$\alpha_1^{(2)}(\x_1,\x_2)=\Phi(\x_1)\Phi(\x_2)$.
It follows that the two-particle density and the 
two-particle momentum distribution factorize up 
to corrections of order ${\mathcal O}(1/N)$
into products of the respective one-particle distributions:
\begin{eqnarray}
\rho^{(2)}_{\vert N,0\rangle}(\x_1,\x_2\vert \x_1,\x_2)	&=& 	N(N-1)\vert\Phi(\x_1)\vert^2\vert\Phi(\x_2)\vert^2\label{rho2xharm},\\
\rho^{(2)}_{\vert N,0\rangle}(\k_1,\k_2\vert \k_1,\k_2)	&=&	N(N-1)\vert\Phi(\k_1)\vert^2\vert\Phi(\k_2)\vert^2\label{rho2kharm}.
\end{eqnarray}
At the barrier height $A=13$, the system is condensed but spread out over the two wells.
Then,  using Eqs.~(\ref{distOrbs}) to approximate $\alpha_1^{(1)}$, we find
\begin{eqnarray}
\rho^{(2)}_{\vert N,0\rangle}(\x_1,\x_2\vert \x_1,\x_2)	&=&\frac{N(N-1)}{4} \left[	
		\vert\Phi_1(\x_1)\Phi_1(\x_2)\vert^2+
		\vert\Phi_1(\x_1)\Phi_2(\x_2)\vert^2+\right.\nonumber\\
& &		\left.\vert\Phi_2(\x_1)\Phi_1(\x_2)\vert^2+
		\vert\Phi_2(\x_1)\Phi_2(\x_2)\vert^2\right],\label{rho2xcond}\\
		\rho^{(2)}_{\vert N,0\rangle}(\k_1,\k_2\vert \k_1,\k_2)	&=& N(N-1)[1+\cos(\k_1d)][1+\cos(\k_2d)]	
		\vert\Phi(\k_1)\Phi(\k_2)\vert^2\label{rho2kcond}
\end{eqnarray}
for the two-particle density and  the two-particle 
momentum distribution. Apart from a correction of order ${\mathcal O}(1/N)$,
the two-particle density and the two-particle momentum distribution are again 
products of the respective one-particle distributions.

\subsubsection{Fully fragmented state}
In Sec.~\ref{fragmentedState} we found that  
three natural geminals are occupied in the 
fully fragmented regime, see Eq.~(\ref{fragGemOcc}).
The occupations of Eq.~(\ref{fragGemOcc})
hold exactly for a state  of the form $\vert N/2,N/2\rangle$. 
Therefore, we approximate the second-order reduced density operator in the fully fragmented 
regime by that of the state $\vert N/2,N/2\rangle$:
\begin{eqnarray}\label{rho2Opfrag}
\hat{\rho}^{(2)}_{\vert N/2,N/2\rangle}     &=& N\frac{N}{2} 
	\vert\alpha^{(2)}_1\rangle\langle\alpha^{(2)}_1\vert+
	\frac{N}{2}\left(\frac{N}{2}-1\right)	\vert\alpha^{(2)}_2\rangle\langle\alpha^{(2)}_2\vert+
	\frac{N}{2}\left(\frac{N}{2}-1\right)   \vert\alpha^{(2)}_3\rangle\langle\alpha^{(2)}_3\vert, 
\end{eqnarray}
where the natural geminals $\vert\alpha^{(2)}_i\rangle$ are given by Eq.~(\ref{fragmentedgems}).
In contrast to the condensed state, the normalized second-order correlation function
of the fully fragmented state has a more complicated structure due to the different 
terms contributing to Eq.~(\ref{rho2Opfrag}).
We approximate the natural  geminals using Eqs.~(\ref{distOrbs})
and find
\begin{eqnarray}
\rho^{(2)}_{\vert N/2,N/2\rangle}(\x_1,\x_2\vert \x_1,\x_2)       &=&
    	\frac{N}{2} \left(\frac{N}{2}-1\right)\left[\vert\Phi_1(\x_1)\Phi_1(\x_2)\vert^2+ 
		\vert\Phi_2(\x_1)\Phi_2(\x_2)\vert^2\right]+\nonumber \\
& &     \frac{N}{2} \frac{N}{2} 
\left[\vert\Phi_1(\x_1)\Phi_2(\x_2)\vert^2+ \vert\Phi_2(\x_1)\Phi_1(\x_2)\vert^2\right],\label{rho2xfrag}\\
\rho^{(2)}_{\vert N/2,N/2\rangle}(k_1,k_2\vert k_1,k_2)   	    &=& 
		N(N-1)\left(1+\frac{N}{N-1}\frac{\cos[(k_1-k_2)d]}{2}\right)\vert\Phi(k_1)\Phi(k_2)\vert^2\label{rho2kfrag}
\end{eqnarray}
for the two-particle density and the two-particle momentum distribution. 
This representation allows us to identify the first 
two terms in Eq.~(\ref{rho2xfrag}) as 
contributions coming from two separate condensates of $N/2$ bosons each, 
with condensate wave functions $\Phi_1(\x_1)$ and $\Phi_2(\x_1)$.
The third term in Eq.~(\ref{rho2xfrag}) 
is due to the fact that the bosons in the two separated 
condensates are identical particles.
For the normalized second-order correlation function one finds:
\begin{eqnarray}\label{g2xMF}
g^{(2)}_{\vert N/2,N/2\rangle}(\x_1,\x_2,\x_1,\x_2)	&=&	
				\begin{cases}
					1-\frac{2}{N}&\text{if } \x_1,\x_2 \in L \text{ or } \x_1,\x_2\in R\\
					1&\text{otherwise }\\
				\end{cases}, 
\end{eqnarray}
which mimics a high degree of second-order coherence.
However, when $g^{(2)}$ is evaluated on the diagonal in momentum space, 
one finds  
\begin{eqnarray}\label{g2kfrag}
g^{(2)}_{\vert N/2,N/2\rangle}(k_1,k_2,k_1,k_2)	&=& 
\left(1-\frac{1}{N}\right)	\left(1+\frac{N}{N-1}\frac{\cos[(k_1-k_2)d]}{2}\right),
\end{eqnarray}
which displays an oscillatory behavior and 
deviates significantly from a uniform 
value of one. Hence the system is clearly not coherent, see Sec.~\ref{secbasic}. 
Therefore, a description of second 
order correlations in terms of 
$\rho^{(2)}(\x_1,\x_2\vert\x_1,\x_2)$ and $g^{(2)}(\x_1,\x_2,\x_1,\x_2)$ is incomplete, and 
$\rho^{(2)}(k_1,k_2\vert k_1,k_2)$ and $g^{(2)}(k_1,k_2,k_1,k_2)$ have to be 
taken into account. Although this may seem obvious in the present 
case of a fully fragmented state, 
this reduction of coherence is more intricate in a 
state which is only partially fragmented, see following subsection.

\subsection{Numerical results}
In this subsection we discuss the second-order correlations 
of the many-body solution. We compare the results to those 
of the BMF solution.
When mean-field theory gives a 
good approximation to the many-body results, 
we also compare with the analytical mean-field model 
of Sec.~\ref{generalResults2}.

\begin{figure}
    \centering
    \includegraphics[width=145mm,angle=0]{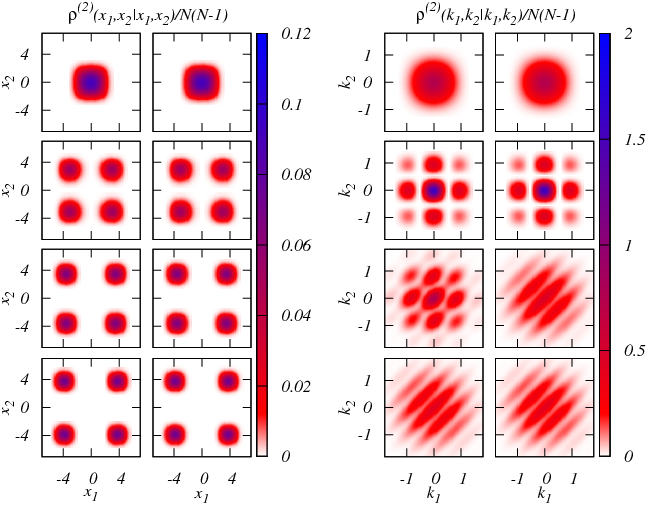}
    	\caption{(color online). Two-particle density and 
		two-particle momentum distribution at different barrier heights.
		The first two columns (from left to right) show the two-particle density 
		$\rho^{(2)}(\x_1,\x_2\vert\x_1,\x_2)/N(N-1)$ of the many-body (left) and BMF (right) wave function 
		for the barrier heights $A=0,13,19,24$, from top to bottom.
		At low barrier heights ($A=0,13$) the system is condensed, 
		and the two-particle RDM factorizes
		into a product of the one-particle RDMs. At higher barriers ($A=19,24$), the system fragments
		and the two-particle RDM does not factorize into a product of the one-particle RDMs. 
		The two-particle density  at high barriers looks very similar 
		to that of the condensed state at $A=13$.
		The fragmentation transition is not visible in the two-particle density. The results of 
		the many-body and BMF wave function
		cannot be distinguished at any barrier height.
		The third and fourth column show the two-particle momentum distribution
		$\rho^{(2)}(\k_1,\k_2\vert \k_1,\k_2)/N(N-1)$ of the many-body  (left) and BMF (right) solution
		at the same barrier heights as above. The transition from a condensed state to a fragmented state
		is clearly visible. At $A=19$ the BMF solution does not reproduce the many-body results. 
		The system is in a true many-body state, inaccessible to mean-field methods.
		At even higher barriers	$A\ge24$ the system fully fragments, 
		and a mean-field description  is applicable again.
		The quantities shown are dimensionless.
		}
    \label{fig5}
\end{figure}

The first two columns of Fig.~\ref{fig5} show 
the two-particle density $\rho^{(2)}(\x_1,\x_2\vert\x_1,\x_2)$
of the many-body (left)  
and BMF (right) solutions at barrier heights 
$A=0,13,19,24$, from top to bottom.
At zero barrier height 
$\rho^{(2)}(\x_1,\x_2\vert\x_1,\x_2)$ 
is localized at the center of the trap. 
The two-particle density factorizes approximately 
into a product of the one-particle densities:
$\rho^{(2)}(\x_1,\x_2\vert\x_1,\x_2)\approx \rho^{(1)}(\x_1\vert\x_1)\rho^{(1)}(\x_2\vert\x_2)$.
This remains true up to barrier heights $A=13$, 
where the condensate is spread out over the two wells.
The BMF result approximates the many-body result well 
in the condensed regime, and the structure of 
$\rho^{(2)}(\x_1,\x_2\vert\x_1,\x_2)$
is that of Eqs.~(\ref{rho2xharm}) and (\ref{rho2xcond}) at barrier heights $A=0$ and $A=13$, respectively.

When the barrier is raised 
further to $A=19$, the system fragments. 
Many permanents contribute to the 
wave function in this regime and there 
is no simple formula that relates 
the occupations of the natural orbitals
to the two-particle density. 
Similar to the one-particle density, described in Sec.~\ref{numericalResults1}, 
the two-particle density seems 
to take no notice of the 
transition from a single to a fragmented condensate. 
It remains practically 
unchanged during the transition, apart from 
a slight shift of the peaks away from each other as the barrier is raised.

At even higher barriers, $A\ge24$, 
the many-body state becomes 
fully fragmented and the wave 
function approaches $\vert N/2,N/2\rangle$. 
In this limit it is 
again possible to describe the 
two-particle density on a mean-field level.
Therefore, the analytical 
results of Sec.~\ref{generalResults2} 
for the fully fragmented state should apply.
In fact, the structure of 
$\rho^{(2)}(\x_1,\x_2\vert\x_1,\x_2)$ 
in the fully fragmented regime
is that predicted by Eq.~(\ref{rho2xfrag}). 

The two-particle density of the 
condensed state
just below the fragmentation  transition
and of the fully fragmented 
state above the fragmentation transition 
cannot be distinguished.
It is easily verified, that Eqs.~(\ref{rho2xcond}) 
and (\ref{rho2xfrag}) give rise to the same 
two-particle density profile up to 
corrections of order ${\mathcal O}(1/N)$.

In contrast, the fragmentation transition 
is clearly visible in the two-particle 
momentum distribution. In the third and 
fourth columns of Fig.~\ref{fig5} the 
two-particle momentum distribution 
$\rho^{(2)}(k_1,k_2\vert k_1,k_2)$ 
of the many-body (left) and  
BMF (right) wave function are shown.

In the condensed regime the two-particle momentum
distribution is approximately given by
the product of one-particle  
momentum distributions of a single condensate. 
This agrees with the analytical
predictions of Eq.~(\ref{rho2kharm}) at barrier height $A=0$
and Eq.~(\ref{rho2kcond}) at $A=13$. 
The mean-field picture is appropriate here. 

In the many-body regime 
the two-particle momentum distribution
$\rho^{(2)}(k_1,k_2\vert k_1,k_2)$  contains
contributions from many permanents.
The resulting $\rho^{(2)}(k_1,k_2\vert k_1,k_2)$ has
a structure that lies somewhat in between 
the two results, Eqs.~(\ref{rho2kcond}) and (\ref{rho2kfrag}), 
obtained within the analytical 
mean-field model.
The BMF solution 
is fully fragmented and 
does not provide an accurate 
approximation to the many-body 
two-particle momentum distribution in this regime, see Fig.~\ref{fig5}, 
third and fourth columns in the third row from above.

When the barrier is raised to $A=24$, 
the many-body state becomes  fully fragmented 
and the mean-field picture is again applicable.
The pattern of a single coherent condensate
has now vanished completely in favor of 
a pattern characteristic of two separate condensates.
The pattern agrees well with the structure 
predicted by Eq.~(\ref{rho2kfrag}).

Similar to our results on first 
order correlations, discussed 
in Sec.~\ref{numericalResults1}, 
the fragmentation
transition shows up in the 
two-particle momentum distribution, 
but not in the two-particle density.
While this behavior is predictable
in the limiting cases of a 
condensed and a fully fragmented
state, it is necessary to solve the many-body
problem to determine the limits 
of such mean-field approximations.
Particularly the behavior in between 
the two mean-field limits 
is only accessible to many-body approaches.

\begin{figure}
    \centering
    \includegraphics[width=145mm,angle=0]{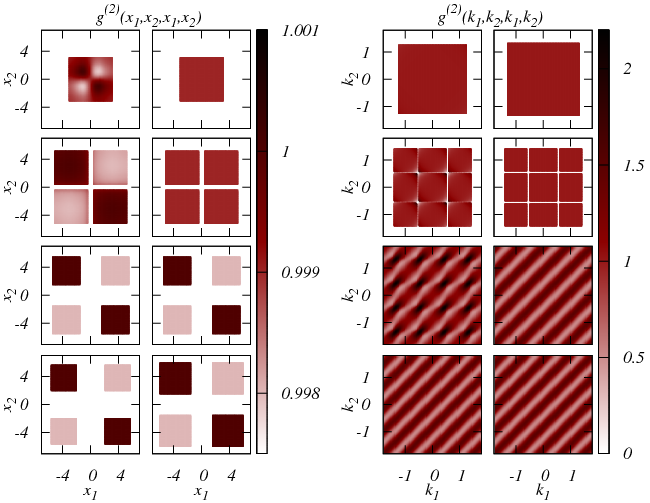}
    	\caption{(color online). Second order coherence at different barrier heights. 
		The first two columns (from left to right) show the diagonal of the 
		normalized second-order correlation function in real space
                $g^{(2)}(\x_1,\x_2,\x_1,\x_2)$ of the many-body (left) and BMF (right) solution
		at barrier heights $A=0,13,19,24$, from top to bottom. 
		$g^{(2)}(\x_1,\x_2,\x_1,\x_2)$ is very close
		to one at all barrier heights. Note the scale! 
		The system seems to be second-order coherent
		and the results of the many-body and BMF solution agree well with each other.
		The third and fourth column depict the diagonal of the
                normalized second-order correlation function in momentum space
                $g^{(2)}(\k_1,\k_2,\k_1,\k_2)$ of the many-body (left) and BMF (right) solution
		at the same barrier heights. The fragmentation transition is clearly visible between 
		$A=13$ and $A=24$. At $A=19$ there are strong many-body correlations between the momenta 
		(local maxima in black color) 	and $g^{(2)}(\k_1,\k_2,\k_1,\k_2)$ exhibits 
		a complicated pattern, see text for more details.
		A mean-field description fails here. 
		In the limit of  high barriers, $A\ge24$, the correlations of the many-body state  
		become again describable by those of the BMF solution.
		The quantities shown are dimensionless.
		}
    \label{fig6}
\end{figure}

We will now address the second-order 
coherence of the system. The first two columns of Fig.~\ref{fig6} 
show the diagonal of the normalized second 
order correlation function 
$g^{(2)}(\x_1,\x_2,\x_1,\x_2)$ 
of the many-body (left) and the 
BMF (right) solutions. Note the scale!  
The Eqs.~(\ref{g2xcondensed}) and (\ref{g2xMF}) 
of the analytical mean-field model of Sec.~\ref{generalResults2} 
predict very small correlations in the two-particle density 
of the condensed and the fragmented state.
This is confirmed in the first column of Fig.~\ref{fig6}.
In the condensed regime at zero barrier height
the effects of the depletion of the condensate
on $g^{(2)}(\x_1,\x_2,\x_1,\x_2)$ are visible. 
Almost no two-particle density correlations are present.
This is equally true in the case of a single condensate
spread out over the two wells and also in the many-body regime. 
Above the fragmentation 
transition, the present computation of 
the many-body solution cannot describe effects on 
$g^{(2)}$ that are due to the 
depletion of the condensate. 
However, since depletion effects 
are negligible in the harmonic
trap, we are reassured that they are 
also negligible in the fully
fragmented regime, see Sec.~\ref{fragmentedState}.
The BMF solution predicts almost identical two-body 
density correlations, see second column of Fig.~\ref{fig6}.

On the basis of $g^{(2)}(\x_1,\x_2,\x_1,\x_2)$ alone, 
the many-body state {\em appears} 
to be second-order coherent
at all barrier heights.  
A high degree of second-order coherence requires Eq.~(\ref{coh}) 
to hold to a very good approximation for $p=1$ and $p=2$. This in turn requires 
the largest eigenvalues of the first- and
second-order RDM to be $n_1^{(1)}\approx N$ 
and $n_1^{(2)}\approx N(N-1)$, respectively.
We have already demonstrated in Sec.~\ref{model} 
that these conditions are only satisfied in the 
condensed regime. Therefore, it is obviously tempting, 
but wrong to conclude from $g^{(2)}(\x_1,\x_2,\x_1,\x_2)\approx1$
that the system is second-order coherent. 
This misconception is due to 
the fact that $g^{(2)}(\x_1,\x_2,\x_1,\x_2)$
only samples a small part of the first 
and second-order RDMs of the system.

So, how does the decrease of coherence manifest itself in second order correlation functions?
For second-order coherence to be present, at least approximately, 
also $g^{(2)}(k_1,k_2,k_1,k_2)$ 
has to be close to one.
The third and fourth column of Fig.~\ref{fig6} 
show $g^{(2)}(k_1,k_2,k_1,k_2)$ 
of the many-body (left) and  BMF (right) solution.
At zero barrier height the system is indeed 
highly second-order coherent since 
only one natural orbital is significantly occupied.
Not only $g^{(2)}(\x_1,\x_2,\x_1,\x_2)$, 
but also $g^{(2)}(k_1,k_2,k_1,k_2)$ is very close to one here. 
However, at $A=13$ 
when the many-body state is still condensed, $g^{(2)}(k_1,k_2,k_1,k_2)$ 
starts to  develop a structure. 

When the barrier is raised to 
values above $A=13$, the structure becomes more 
and more pronounced.
In the many-body regime at $A=19$, we find that 
$g^{(2)}(k_1,k_2,k_1,k_2)$ has a complicated 
behavior and deviates significantly from values close 
to one, thereby proving that strong correlations are present.
Note that the interaction between the particles is weak and that
the strong correlations are due to the transition 
from a single to a fragmented condensate. 
This transition is in turn induced by a change of the 
shape of external potential.
Varying the shape of the external potential therefore provides 
a means to introduce strong correlations between the particles.
The strongest correlations 
(black spots in the third panel of the third row of Fig.~\ref{fig6}) 
occur at those values where the two-body momentum distribution
has local minima. At the values of $\k_1$ and $\k_2$, where the strongest 
correlations occur, the one-body and the 
two-body momentum distributions are {\em clearly} 
distinct from zero, see third panel in the middle column of  Fig.~\ref{fig3} 
and the third panel in the third row of Fig.~\ref{fig5}.  
Experiments that measure $g^{(2)}(k_1,k_2,k_1,k_2)$ in ultracold quantum gases
have been carried out recently, see e.g. \cite{PCK07}.
An experiment that measures 
$g^{(2)}(k_1,k_2,k_1,k_2)$ 
would find  strong two-particle momentum correlations 
at high barriers.

When the system becomes fully fragmented
at barrier heights $A\ge24$ the structure of 
$g^{(2)}(k_1,k_2,k_1,k_2)$ becomes more regular again. 
The amplitude of the correlations 
is smaller than in the many-body regime, and 
the correlations between different momenta are 
modulated by a single oscillatory structure.
This structure can be well understood 
within the analytical mean-field model
of Sec.~\ref{generalResults2}.
The oscillatory part of $g^{(2)}(k_1,k_2,k_1,k_2)$
is determined by the difference of the wave
vectors multiplied by the distance between the wells, 
see Eq.~(\ref{g2kfrag}). 

Hence, we find that only in the condensed regime 
the system is second-order coherent
despite the fact 
that $g^{(2)}(\x_1,\x_2,\x_1,\x_2)\approx 1$ at 
all barrier heights. This merely reflects 
the fact that $g^{(2)}(\x_1,\x_2,\x_1,\x_2)$ is only the diagonal
of $g^{(2)}(\x'_1,\x'_2,\x_1,\x_2)$. 
On the other hand, $g^{(2)}(k_1,k_2,k_1,k_2)$ 
depends on all values of 
of $\rho^{(2)}(\x_1,\x_2\vert\x'_1,\x'_2)$ and 
provides complementary information about the coherence of the state. 
A description of second-order coherence
in terms of $g^{(2)}(\x_1,\x_2,\x_1,\x_2)$ alone 
is therefore incomplete. 

The corresponding results of the 
BMF solution agree well with those of 
the many-body solution as long as 
the system is not in the many-body regime at intermediate barrier heights. 
In the many-body regime the BMF result is inaccurate, but 
it anticipates the final form of $g^{(2)}(\x_1,\x_2,\x_1,\x_2)$ 
in the fragmented regime.

\section{CONCLUSIONS}
In this work we have investigated first- and second-order 
correlations of trapped interacting bosons. 
For illustration purposes we have 
investigated the ground state 
of $N=1000$ weakly interacting 
bosons in a one-dimensional double-well trap
geometry at various barrier heights on a many-body level. 
We have obtained the many-body results 
by solving the many-body Schr\"odinger equation
with the recently developed MCTDHB method. 
This allowed us to compute from first principles
the natural orbitals and the natural geminals of 
a large interacting many-body system, 
together with their occupation numbers.
To our knowledge this is the first 
computation of the natural geminals
of an interacting many-body system of this size.

Depending on the height of the double-well barrier
we found that there are three different parameter regimes.
At low barriers the ground state is condensed 
and the many-body wave function
is well approximated by a single permanent of the form $\vert N,0\rangle$.
At high barriers the ground state becomes fully fragmented 
and can be well approximated 
by a single permanent of the form $\vert N/2,N/2\rangle$.
At intermediate barrier heights, 
where the transition from a single to a 
fragmented condensate occurs, 
the ground state
becomes a true many-body wave function 
to which many permanents contribute.   
We have demonstrated that the transition to a fragmented state 
results in the occupation of a natural geminal that
contributes very little  to the interaction energy.
The overall energy of the system can be lowered 
by the occupation of such a geminal, and 
the ground state becomes fragmented.

We have shown how the transition from a condensed to 
a fully fragmented ground state manifests itself in
the one- and two-particle momentum distributions.
However, the transition is {\em not}
captured by the one- and 
two-particle density distributions, 
not even in the many-body regime.

In order to determine the coherence of the 
state during the fragmentation transition, 
we have computed the first- and second-order 
normalized correlation functions $g^{(1)}(\x_1',\x_1)$, 
$g^{(2)}(\x_1,\x_2,\x_1,\x_2)$ and $g^{(2)}(\k_1,\k_2,\k_1,\k_2)$. 
In the condensed regime, a high degree of coherence 
is indeed present in the ground state wave function. 
First and second order correlations were 
found to be negligible 
at the interaction strength and particle number 
chosen for our computation.
However, with increasing barrier height
correlations between the momenta 
of the particles build up. These correlations 
were found to be very strong in the many-body regime at 
intermediate barrier heights.
The ground state at high barriers was found to
be correlated, but not as strongly as the ground state
at intermediate barrier heights.

While the transition from a virtually
uncorrelated state to a correlated one 
is clearly visible in $g^{(1)}(\x_1',\x_1)$
and $g^{(2)}(\k_1,\k_2,\k_1,\k_2)$, 
the transition hardly shows up 
in $g^{(2)}(\x_1,\x_2,\x_1,\x_2)$. 
A description of second-order coherence
in terms of $g^{(2)}(\x_1,\x_2,\x_1,\x_2)$ 
alone is, therefore,  incomplete
and can lead to wrong predictions.

For comparison we have computed results based on (i)
the best approximation of the many-body wave function
within mean-field theory, the BMF wave function, and (ii) 
the Gross-Pitaevskii solution. 
We found that the GP wave function is identical to
the BMF solution up to some barrier height.
However, once the true many-body solution starts to fragment
the BMF wave function is no longer given by  a GP type 
permanent $\vert N,0\rangle$, but rather by a fragmented
state of the form $\vert N/2,N/2\rangle$.
In the true many-body regime neither the GP, nor the BMF solution 
provide an adequate approximation to the many-body wave function,
and the predicted correlations are inaccurate.

While the GP mean-field is only accurate  
at low barrier heights, 
the BMF solution provides a very good approximation to 
the true many-body wave function at 
low {\em and} high barriers.
We have shown that the GP mean-field 
predicts qualitatively wrong results at high barriers.
The BMF only fails at intermediate 
barrier heights where the true many-body 
wave function becomes a 
superposition of many permanents. 
Such many-body effects can, by construction, 
not be captured by mean-field methods. 

In the mean-field regimes at high and low barriers 
we have provided an analytical mean-field model
that allows us to understand the general structure
of the computed correlation functions. 

Our work sheds new 
light on the first- and second-order 
correlation functions 
of interacting many-body systems. 
The variation of 
the shape of the 
trapping potential 
allows one to change the physics of the system 
from mean-field to strongly correlated
many-body physics.
Particularly, the many-body regime
in between the condensed and the 
fully fragmented regimes has shown to 
be very rich and promises exciting results for experiments to come.

\section*{ACKNOWLEDGMENTS}
Financial support by the DIP - Deutsch-Israelische 
Projektkooperation and the DFG - Deutsche Forschungsgemeinschaft
is gratefully acknowledged.

\appendix
\section{$p$-particle momentum distribution}\label{A}
The relation of the $p$-particle momentum distribution to the $p$-th order RDM is shown.
The $p$-particle RDM is related 
to the $p$-particle momentum distribution by
\begin{equation}\label{momA}
\rho^{(p)}(\p_1,\dots,\p_p\vert\p_1,\dots,\p_p;t)
=\frac{1}{(2\pi)^{Dp}}\int d^p\r d^p\r'\, e^{-i\sum_{l=1}^p \p_l(\r_l-\r'_l)} 
\rho^{(p)}(\r_1,\dots,\r_p\vert\r'_1,\dots,\r'_p;t).
\end{equation}
The change of variables $\R_i=\frac{\r_i+\r'_i}{2}$, $\s_i=\r_i-\r'_i$ for $i=1,\dots,p$ 
in Eq.~(\ref{momA}) leads to
\begin{eqnarray}\label{rhopgamma}
\rho^{(p)}(\p_1,\dots,\p_p\vert\p_1,\dots,\p_p;t)
&=&\int d^p\s \, e^{-i\sum_{l=1}^p \p_l\s_l}\gamma^{(p)}(\s_1,\dots,\s_p;t),
\end{eqnarray}
where
\begin{equation}\label{gammaA}
\gamma^{(p)}(\s_1,\dots,\s_p;t)
=\int d^p\R\, \rho^{(p)}(\R_1+\frac{\s_1}{2},\dots,\R_p+\frac{\s_p}{2}\vert\R_1-\frac{\s_1}{2},\dots,\R_p-\frac{\s_p}{2};t)
\end{equation}
is the average of the value of $\rho^{(p)}(\r_1,\dots,\r_p\vert \r'_1,\dots,\r'_p;t)$ at distances $\s_i$ between
$\r_i$ and $\r'_i$. From Eqs.~(\ref{rhopgamma}-\ref{gammaA}) 
it is clear that the $p$-particle momentum distribution 
at large momenta is determined by the behavior of 
$\rho^{(p)}(\r_1,\dots,\r_p\vert \r'_1,\dots,\r'_p;t)$  
at short distances, whereas at low momenta the off-diagonal long range behavior 
of $\rho^{(p)}(\r_1,\dots,\r_p\vert \r'_1,\dots,\r'_p;t)$ contributes the major part.

\clearpage

\begin{thebibliography}{99}

\bibitem{firstBEC1} M. H. Anderson, J. R. Ensher, M. R. Matthews, C. E. Wieman, and E. A. Cornell, 
			Science {\bf 269}, 198 (1995). 

\bibitem{firstBEC2} C. C. Bradley, C. A. Sackett, J. J. Tollett, and R. G. Hulet, Phys. Rev. Lett. {\bf 75}, 1687 (1995).
\bibitem{firstBEC3} K. B. Davis, M.-O. Mewes, M. R. Andrews, N. J. van Druten, D. S. Durfee, 
	D. M. Kurn, and W. Ketterle, Phys. Rev. Lett. {\bf 75}, 3969 (1995).
\bibitem{BGM97} E. A. Burt, R. W. Ghrist, C. J. Myatt, M. J. Holland, E. A. Cornell, and C. E. Wieman, Phys. Rev. Lett. {\bf 79}, 337 (1997).
\bibitem{CHK03} L. Cacciapuoti, D. Hellweg, M. Kottke, T. Schulte, W. Ertmer, J. J. Arlt, K. Sengstock, L. Santos, and M. Lewenstein, Phys. Rev. A {\bf 68}, 053612 (2003).
\bibitem{LHP04} B. L. Tolra, K. M. O'Hara, J. H. Huckans, W. D. Phillips, S. L. Rolston, and J. V. Porto, Phys. Rev. Lett. {\bf 92}, 190401 (2004).
\bibitem{FGW05} S. F\"olling, F. Gerbier, A. Widera, O. Mandel, T. Gericke and I. Bloch, Nature {\bf 434}, 481 (2005). 
\bibitem{SHP05} M. Schellekens, R. Hoppeler, A. Perrin, J. V. Gomes, D. Boiron, A. Aspect, and C. I. Westbrook, Science {\bf 310}, 648 (2005).
\bibitem{ORK05} A. \"Ottl, S. Ritter, M. K\"ohl, and T. Esslinger, Phys. Rev. Lett. {\bf 95}, 090404 (2005).
\bibitem{SHL07} L. E. Sadler, J. M. Higbie, S. R. Leslie, M. Vengalattore, and D. M. Stamper-Kurn, Phys. Rev. Lett. {\bf 98}, 110401  (2007). 
\bibitem{ROD07} S. Ritter, A. \"Ottl, T. Donner, T. Bourdel, M. K\"ohl, and T. Esslinger, Phys. Rev. Lett. {\bf 98}, 090402 (2007).
\bibitem{NaG99} M. Naraschewski and R. J. Glauber, Phys. Rev. A {\bf 59}, 4595 (1999).
\bibitem{GPS06}	J. V. Gomes, A. Perrin, M. Schellekens, D. Boiron, C. I. Westbrook, and M. Belsley, Phys. Rev. A {\bf 74}, 053607 (2006).
\bibitem{HoC99} M. Holzmann and Y. Castin, Eur. Phys. J. D {\bf 7}, 425 (1999).	
\bibitem{KGD03} K. V. Kheruntsyan, D. M. Gangardt, P. D. Drummond, and G. V. Shlyapnikov, Phys. Rev. Lett. {\bf 91}, 040403 (2003).

\bibitem{AsG03} G. E. Astrakharchik and S. Giorgini, Phys. Rev. A {\bf 68}, 031602 (2003). 
\bibitem{BaR04} R. Bach and K. Rz\c{a}\.{z}ewski, Phys. Rev. A {\bf 70}, 063622 (2004).
\bibitem{PeB07} R. Pezer and H. Buljan, Phys. Rev. Lett. {\bf 98}, 240403 (2007). 
\bibitem{ZMS06a} S. Z\"ollner, H.-D. Meyer, and P. Schmelcher, Phys. Rev. A {\bf 74}, 053612 (2006).
\bibitem{ZMS06b} S. Z\"ollner, H.-D. Meyer, and P. Schmelcher, Phys. Rev. A {\bf 74}, 063611 (2006).
\bibitem{ZMS08}  S. Z\"ollner, H.-D. Meyer, and P. Schmelcher, Phys. Rev. Lett. {\bf 100}, 040401 (2008).
\bibitem{PeO56} O. Penrose and L. Onsager, Phys. Rev. {\bf 104}, 576 (1956).
\bibitem{NoS82} P. Nozi\`eres, D. Saint James, J. Phys. (Paris) {\bf 43}, 1133 (1982);
P. Nozi\`eres, in {\em Bose-Einstein Condensation}, edited by A. Griffin, D. W. Snoke, and S. Stringari (Cambridge University Press, Cambridge, England, 1996).
\bibitem{SpS99} R. W. Spekkens and J. E. Sipe, Phys. Rev. A {\bf 59},  3868 (1999).
\bibitem{CeS03} L. S. Cederbaum and A. I. Streltsov, Phys. Lett. A {\bf 318}, 564 (2003).
\bibitem{SCM04} A. I. Streltsov, L. S. Cederbaum, and N. Moiseyev, Phys. Rev. A {\bf 70}, 053607 (2004).


\bibitem{AlC05} O. E. Alon and L. S. Cederbaum, Phys. Rev. Lett. {\bf 95}, 140402 (2005).



\bibitem{ASC05b} O. E. Alon, A. I. Streltsov, and L. S. Cederbaum, Phys. Lett. A {\bf 347}, 88 (2005).
\bibitem{SAC06} A. I. Streltsov, O. E. Alon, and L. S. Cederbaum, Phys. Rev. A {\bf 73}, 063626 (2006).

\bibitem{SAC07} A. I. Streltsov, O. E. Alon, and L. S. Cederbaum, Phys. Rev. Lett. {\bf 99}, 030402 (2007).
\bibitem{ASC07b} O. E. Alon, A. I. Streltsov, and L. S. Cederbaum, J. Chem. Phys. {\bf 127}, 154103 (2007). 
\bibitem{ASC07c} O. E. Alon, A. I. Streltsov, and L. S. Cederbaum, cond-mat/0703237, PRA in press.


\bibitem{ASC05a} O. E. Alon, A. I. Streltsov, and L. S. Cederbaum, Phys. Rev. Lett. {\bf 95}, 030405 (2005).

\bibitem{CoY00} A. J. Coleman, V. I. Yukalov, {\em Reduced Density Matrices: Coulson's Challenge} (Springer, Berlin, 2000).
\bibitem{Maz07} D. A. Mazziotti, Ed., {\em  Reduced-Density-Matrix Mechanics: 
				With Application to Many-Electron Atoms and Molecules}
				(Advances in Chemical Physics), Vol. 134 (Wiley, New York, 2007).
\bibitem{Yan62} C. N. Yang, Rev. Mod. Phys. {\bf 34}, 694 (1962).
\bibitem{Sas65} F. Sasaki, Phys. Rev. {\bf 138}, B1338 (1965).
\bibitem{Gla63} R. J. Glauber, Phys. Rev. {\bf 130}, 2529 (1963).
\bibitem{TiG65} U. M. Titulaer and R. J. Glauber, Phys. Rev. {\bf 140}, B676 (1965).
\bibitem{BHE00} I. Bloch, T. W. H\"ansch, and T. Esslinger, Nature {\bf 403}, 166 (2000).
\bibitem{XSM06} H. Xiong, S. Liu, and M. Zhan, New J. Phys. {\bf 8}, 245 (2006).
\bibitem{CSB06} L. S. Cederbaum, A. I. Streltsov,  Y. B. Band, and O. E. Alon, cond-mat/0607556.
\bibitem{CSB07} L. S. Cederbaum, A. I. Streltsov, Y. B. Band, and O. E. Alon, Phys. Rev. Lett. {\bf 98}, 110405 (2007).
\bibitem{Gir60} M. Girardeau, J. Math. Phys. {\bf 1}, 516 (1960).
\bibitem{LiL63} E. H. Lieb and W. Liniger, Phys. Rev. {\bf 130}, 1605 (1963).
\bibitem{McG64} J. B. McGuire, J. Math. Phys. {\bf 5}, 622 (1964).
\bibitem{GiW00} M. D. Girardeau and E. M. Wright, Phys. Rev. Lett. {\bf 84}, 5691 (2000).
\bibitem{YuG05} V. I. Yukalov and M. D. Girardeau, Laser Phys. Lett. {\bf 2}, 375 (2005).
\bibitem{SSA05} K. Sakmann, A. I. Streltsov, O. E. Alon, and L. S. Cederbaum, Phys. Rev. A {\bf 72}, 033613 (2005).
\bibitem{Gir06} M. D. Girardeau, Phys. Rev. Lett. {\bf 97}, 100402 (2006).
\bibitem{SDD07} A. G. Sykes, P. D. Drummond, and M. J. Davis, Phys. Rev. A {\bf 76}, 063620 (2007).
\bibitem{BPG07} H. Buljan, R. Pezer, and T. Gasenzer, arXiv:0709.1444, PRL in press.
\bibitem{Gro61} E. P. Gross, Nuovo Cimento {\bf 20}, 454 (1961).
\bibitem{Pit61} L. P. Pitaevskii, Zh. Eksp. Teor. Fiz. {\bf 40}, 646 (1961); Sov. Phys. JETP {\bf 13}, 451 (1961).
\bibitem{PeS02} C. J. Pethick and H. Smith, {\em Bose-Einstein Condensation in Dilute Gases} (Cambridge University Press, Cambridge, England, 2002).



\bibitem{PiS99} L. Pitaevskii and S. Stringari, Phys. Rev. Lett. {\bf 83}, 4237 (1999).
\bibitem{PCK07} A. Perrin, H. Chang, V. Krachmalnicoff, M. Schellekens, D. Boiron, A. Aspect, and C. I. Westbrook, Phys. Rev. Lett. {\bf 99}, 150405 (2007).

\end{thebibliography}
\end{document}